\documentclass[12pt]{article}
\usepackage[dvips]{graphicx}
\usepackage{epsfig}
\usepackage{amsmath}
\usepackage{bm}
\usepackage{times}
\usepackage{psfrag}
\makeatletter
\@addtoreset{equation}{section}
\renewcommand\theequation{\thesection.\@arabic\c@equation}
\makeatother

\newcommand{\calo}{{\cal O}}

\newcommand{\mtaud} {m^2_{\tau}}
\newcommand{\mkd} {m^2_K}

\newcommand{\mpid} {m^2_\pi}

\def\roughly#1{\mathrel{\raise.3ex\hbox{$#1$\kern-.75em%
\lower1ex\hbox{$\sim$}}}}

\begin{document}

\begin{titlepage}

\begin{center}
{\LARGE\bf Dispersive representation and shape of the $K_{l3}$ form factors:
 robustness.}\\[12mm]

{\normalsize\bf V\'eronique Bernard~${}^{a,}$\footnote {Email:~bernard@ipno.in2p3.fr},
Micaela Oertel~${}^{b,}$\footnote {Email:~micaela.oertel@obspm.fr},
Emilie Passemar~${}^{c,}$\footnote{Email:~passemar@itp.unibe.ch} and Jan Stern~${}^{a,}$
\footnote{Jan Stern sadly passed away before the final version of the paper was written.}} \\[4mm]

{\small\sl ${}^{a}$ Groupe de Physique Th\'{e}orique, IPN,
         CNRS/Univ. Paris-Sud 11, 91406 Orsay Cedex, France}\\
\hspace{-0.3cm}{\small\sl ${}^{b}$ LUTH, Observatoire de Paris, CNRS, Univ.
  Paris Diderot, 5 place Jules Janssen, 92195 Meudon, ~France} \\
{\small\sl ${}^{c}$ Institute for theoretical physics,
University of Bern, Sidlerstr. 5, 3012 Bern, Switzerland}\\
[12mm]
\end{center}

\noindent{\bf Abstract:}
{An accurate low-energy dispersive parametrization of the 
scalar $K \pi$ form factor was constructed some time ago in terms of a single 
parameter guided by the Callan-Treiman low-energy theorem. A similar 
twice subtracted dispersive parametrization 
for the vector $K \pi$ form factor will be investigated here. The robustness
of the parametrization of these two form factors will be studied in great 
detail. In particular the cut-off dependence, the isospin breaking effects and
the possible, though not highly probable, presence of zeros in
the form factors will be discussed. Interesting constraints in the latter case 
will be obtained from the soft-kaon analog of the Callan-Treiman theorem and 
 a comparison with the recent $\tau \rightarrow K \pi \nu_\tau $ data.}

\noindent
\end{titlepage}
\pagenumbering{arabic}
\renewcommand{\thefootnote}{\arabic{footnote}}
\parskip12pt plus 1pt minus 1pt
\topsep0pt plus 1pt
\setcounter{totalnumber}{12}
\section{Introduction}
Experimental information on the shape of the strangeness
changing scalar $f_0(t)$ and vector $f_+(t)$ form factors in the 
low-energy region can be obtained from the study of $K_{\ell3}$-decays. 
$f_0(t)$ and  $f_+(t)$ indeed  enter the differential decay rates of these semi-leptonic
processes. In the expression of these decay rates $f_0(t)$ is multiplied by a 
kinematic factor $(m_\ell/m_K)^2$ with $m_\ell$ and $m_K$ the 
lepton and the kaon mass, respectively. This factor being of the order $10^{-6}$ 
for the electron, only the muon mode is, in fact,
sensitive to the scalar form factor which is thus harder to
determine. Different collaborations, namely ISTRA \cite{ISTRA}, 
KLOE \cite{KLOEe,KLOEmu}, KTeV \cite{KTeVe,KTeVz} and NA48 \cite{NA48e,NA48mu}
have 
measured 
these kaon  decays. 
In the analysis of their data, they 
parametrize the two form factors in terms of some 
free parameters. 
The actual number of parameters which can be determined from a fit
to the data are,  
due to strong
correlations between them, 
 at most two
for the
vector form factor and one for the scalar one.
Thus quadratic-, 
pole- and more recently parametrizations
based on conformal mapping (denoted z-parametrization) have been used
for $f_+(t)$ while, up to recently $f_0(t)$ was described in terms of
a linear and a pole one~{\footnote {Note that in the case of the 
scalar form factor the lightest resonance, the $\kappa^*$, is 
rather broad. In addition, there is a second resonance not far away. 
Thus contrary to the vector form factor case, the parametrization of the 
scalar form factor in terms of one 
real pole has no physical motivation.}.
While the pole parametrization gives comparable results for $f_+(t)$
in the different experiments, the situation for $f_0$ was much more
confused.  The slope of the normalized scalar form factor,
$\bar{f}_0(t)$, varied typically between $9 \cdot 10^{-3}$ and $15
\cdot 10^{-3}$ depending on the experiments \cite{flavianet}.  Note
that the slope determined in this way can only be an upper limit of
the true mathematical slope $d \bar{f}_0(t)/ dt |_{t=0}$ as calculated
for example within chiral perturbation theory \cite{GL84, KN08}.
Therefore it seemed appropriate, in particular in the case of the
scalar form factor, to develop another parametrization which is as
model independent as possible, involves only one parameter and
determines the higher order terms in the series' expansion on physical
ground.  An accurate dispersive representation has been constructed in
Ref. \cite{Bernard:06}  which fulfills all these properties. Two classes of
parametrizations can hence be distinguished depending on whether or
not physical information is used. In the first class one finds for
example the pole one for the vector form factor and this dispersive
parametrization for the scalar form factor and in the second one the
linear and the quadratic parametrizations.  Strictly speaking the
z-parametrization enters also this latter class. However, in
Ref. \cite{hill} it was shown that under certain conditions
it is 
possible to impose a bound on  the sum of the expansion 
coefficients based on unitarity and the total rate of 
$\tau \to K \pi \nu_\tau$.

While the main aim of these $K_{\ell3}$ experiments was to extract the 
CKM matrix element $V_{us}$ 
the dispersive parametrization of $\bar{f}_0$ 
provides another test of the Standard Model (SM) through
the measurement of
the only unknown parameter $\ln C$, with $C$ the value of the scalar
form factor at the Callan Treiman (CT) point. This was in fact
the original idea behind writing such a dispersive parametrization. 
It has first been used by the NA48
collaboration leading to a rather small slope for $\bar{f}_0$ and a
$4.5 \sigma$ deviation to the SM \cite{NA48mu}.  Unfortunately the
situation is still unclear for the scalar form factor, recent
determinations of $\ln C$ from KLOE \cite{KLOEmu} and KTeV
\cite{KTeVmu} lead to no/slight discrepancy with the SM. Due to the
importance of such a measurement it is very important to discuss in
more detail the robustness of this dispersive parametrization. This is
the main aim of this paper together with the investigation of a
similar parametrization for the vector form factor improving on the pole
parametrization.
     
After introducing basic notations and properties in
section~\ref{sec:definitions} we will briefly  review in
section~\ref{sec:dispersive} the dispersive parametrization for the
scalar form factor. We will then describe an analogous parametrization
for the vector form factor.  In section~\ref{sec:error} we will
investigate the robustness of these parametrizations for both, the
scalar and the vector form factors.  We will discuss the uncertainties
due to the input parameters as well as the expected size of isospin
breaking effects. 
In the dispersive parametrization the standard, since most likely, hypothesis of the
absence of zeros in the form factor has been made. 
In section~\ref{Sec:zero} we will question this hypothesis and
in particular, we will study the presence
of possible real or complex zeros in the form factors and the impact
of these zeros on the value of $\ln C$. Even if the likelihood of such
a scenario is small its study is required by the particular
importance of an unambiguous test of the SM.  The possibility of
discarding zeros in the form factors from some properties of the
scalar form factor as well as from a comparison with high energy data
from $\tau$-decays will be discussed in detail. We will conclude in
section~\ref{sec:summary} and finally present some useful expressions
to simplify the use of the dispersive parametrization in the data 
analysis in the appendix.
\section{Basic definitions and properties}
\label{sec:definitions}
The hadronic matrix element describing       
 $K_{\ell3}$-decays is written in terms of two form factors $f^{K \pi}_+(t)$
and $f^{K \pi}_-(t)$,
\begin{equation}
\langle \pi(p_\pi) | \bar{s}\gamma_{\mu}u | K(p_K)\rangle = 
(p_\pi+p_K)_\mu\  f^{K \pi}_+ (t) + (p_K-p_\pi)_\mu\  f_-^{K \pi } (t)~,         
\label{hadronic element}
\end{equation}
where $t=(p_K-p_\pi)^2$.
The vector form factor $f^{K \pi}_+ (t)$ represents
the P-wave projection of the crossed channel matrix element
$\langle 0 |\bar{s}\gamma_{\mu}u | K\pi \rangle$, whereas                
the S-wave projection is described by the scalar form factor defined as
\begin{equation}
f_0 (t) = f^{K \pi}_+ (t) + \frac{t}{m^2_{K } - m^2_{\pi }} f^{K \pi}_-(t).
\label{defffactor}
\end{equation}
In the following discussion we will  consider the normalized form factors    
\begin{equation}
\bar{f}_0(t)=\frac{f_0(t)}{f_+(0)}~~\mathrm{and}~~\bar{f}_+(t)=\frac{f_+(t)}{f_+(0)}~~\mathrm{with}~~
\bar{f}_0(0)=\bar{f}_+(0)= 1~,
\label{defnffactor}
\end{equation}
and try to describe their shape as precisely as possible in the physical region
of $K_{\ell3}$-decays,  
\begin{figure}[h!!]
\begin{center}
\vskip 0.6cm 
\psfrag{x}{$m_\ell^2$}
\psfrag{xx}{$t_0$}
\psfrag{xxx}{$\Delta_{K \pi}$}
\psfrag{xxxx}{$t_{K \pi}$}
\psfrag{xxxxx}{$K_{\ell 3}$}
\psfrag{xxxxxx}{$K \pi$}
\psfrag{yyyy}{$t[GeV^2]$}
\includegraphics*[scale=0.5]{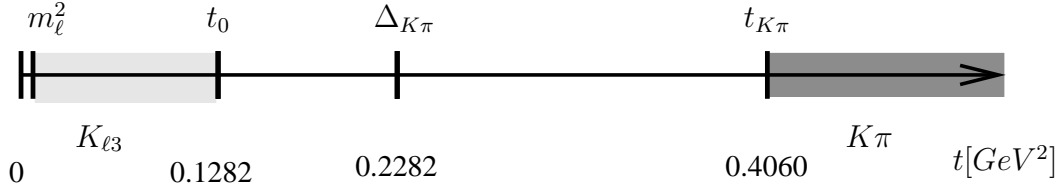}
\vskip 0.4cm
\caption{\it Different energy scales involved in the analysis of the scalar and vector form 
factors: the physical region of $K_{\ell 3}$-decays lies between the 
lepton mass squared $m_\ell^2$ and $t_0=(m_K-m_\pi)^2$, $ \Delta_{K \pi}=m_K^2-m_\pi^2$ 
denotes the CT point and the right-hand cut 
from $K \pi$ scattering starts at $t_{K \pi} =(m_K+m_\pi)^2$. For the numerical values, 
$m_{K^+}$ and $m_{\pi^0}$ have been used.
\label{Cinematic1}}
\end{center}
\end{figure}
$m_\ell^2 \leq t \leq t_0=(m_K-m_\pi)^2$, with $m_\pi$ the pion mass. 
It is shown in Fig.~\ref{Cinematic1} together with the right hand cut
from $K \pi$ scattering which starts at $t_{K \pi} =(m_K + m_\pi)^2$ as
well as the CT point $\Delta_{K \pi}=m_K^2-m_\pi^2$ whose value is
about twice as large as $t_0$. This point is of special interest in
the case of the scalar form factor. Indeed, the Callan-Treiman
low-energy theorem \cite{Dashen:1969bh} predicts its value in the
$SU(2)$ chiral limit (where the quark masses $m_{u,d}$ vanish) at that
particular point. As we will see, this is of great importance in
testing the SM.  For physical quark masses, one has
\begin{equation}
C\equiv
\bar{f}_0(\Delta_{K\pi})=\frac{F_{K^+}}{F_{\pi^+}}\frac{1}{f_{+}^{K^0\pi^-}(0)}+
\Delta_{CT}~,
\label{C} 
\end{equation}
where $F_{K^+,\pi^+}$ are the charged kaon and pion decay constants, respectively,
and $\Delta_{CT}$ is a correction of $ \calo \left( m_{u,d}/4 \pi 
F_\pi \right)$ {\footnote{Note that for practical purposes we have introduced 
some isospin breaking effects already in the first term on the right-hand side (RHS) of Eq.~(\ref {C}).}}.  
It has been estimated within Chiral Perturbation
Theory (ChPT) at next to leading order (NLO) in the isospin
limit~\cite{Gasser:1984ux} with the result
\begin{equation} 
\Delta^{NLO}_{CT}=(-3.5 \pm 8) \times 10^{-3}~.
\label{Delta_CT}
\end{equation}
The error is a conservative estimate assuming typical size
corrections of $ \calo \left( m_{u,d} \right)$ and $ \calo \left( m_s
\right)$ \cite{Leutwyler:07}  for the higher orders. It should 
certainly hold for the neutral 
kaon decays which we are mainly interested in at present. Indeed, 
no large corrections to this estimate are expected
due to the absence of $\pi^0$-$\eta$ mixing in the final state which could
lead to small energy denominators. 
Eq.~(\ref {Delta_CT}) is consistent with the values obtained recently, see Refs~\cite{KN08},
~\cite{Bijnens:07} 
and~\cite{BP08}.
The quantities entering the expression of $C$, Eq. (\ref{C}), are in
principle completely determined within QCD.  Except for
$\Delta_{CT}$, they are studied within lattice QCD, see Ref. \cite{Lellouch} for a recent 
overview of the situation. However, 
the actual most precise determinations of these quantities are obtained from 
semi-leptonic decays. Consequently, they depend on the assumption made 
for the electroweak couplings of quarks.
Assuming the SM couplings, one can extract the quantity
\begin{equation}
B_{exp}=\left|\frac{F_{K^+} V_{us}}{F_{\pi^+} V_{ud}}\right|
\frac{1}{|f_+^{K^0\pi^+}(0) V_{us}|}|V_{ud}|~,  
\label{CGEN}
\end{equation}
using experimental information on the ratio $\Gamma_{K^+_{\ell
2(\gamma)}}/\Gamma_{\pi^+_{\ell 2 (\gamma)}}$, 
the decay $K^0\to \pi^- e \nu_e$~\cite{BM} and $0^+ \to 0^+$
transitions in nuclei~\cite{Towner:08}. These determine respectively
the first ratio, the second one and $|V_{ud}|$ in Eq. (\ref{CGEN})  
with high precision and lead to
\begin{equation}
B_{exp}=1.2418 \pm 0.0039~.
\label{Bexp}
\end{equation}
Interestingly the knowledge of $V_{us}$ is unnecessary for determining $B_{exp}$. 
Eq.~(\ref{C}) then becomes
\begin{eqnarray}
\ln C|_{SM} &=&\ln B_{exp} + \Delta_{CT}/B_{exp} \nonumber \\ 
&=& 0.2166 \pm 0.0034 + 
(-0.0035 \pm 0.0080)/( 1.2418 \pm 0.0039) \nonumber \\
&=& 0.2138 \pm 0.0073\, ,
\label{lnC}
\end{eqnarray}
where in the last line all the errors have been added in quadrature.
Note that the error on $\ln C|_{SM}$ is rather small. 
Hence a precise measurement of
$\ln C$ in neutral $K_{\mu3}$-decays should allow to test the SM
electroweak couplings 
by comparing the obtained value with the one from  Eq.~(\ref{lnC}). 
However, the Callan-Treiman point is
unreachable by a direct measurement of this decay, its value being
much larger than the end point value of the physical region, see
figure~\ref{Cinematic1}. For this reason a dispersive representation
of the scalar form factor written in terms of $\ln C$ as the only 
free parameter has been
introduced in Ref. \cite{Bernard:06}.

\section{Dispersive parametrization}
\label{sec:dispersive}
Let us first briefly review this dispersive parametrization of the scalar
form factor based on an Omn\`es representation \cite{omnes}, 
see also Ref. \cite{Donoghue} for an early application. 
We will then introduce a similar representation to accurately describe 
the vector form factor in the low energy region.

\subsection{Scalar form factor}
\label{sec:scalar}
The dispersive representation of the scalar form factor introduced in
Ref.~\cite{Bernard:06} follows previous attempts 
to determine the form
factor in the physical region. In Ref. \cite{jop00} a coupled channel
approach was used with only one subtraction in the dispersion 
relation.  
The main aim of Ref.~\cite{Bernard:06} was to stay as model
independent as possible. Therefore a second subtraction has been made 
in order to
minimize the bad knowledge of the high energy region in the dispersive
integral. Using the two points $t=0$ (where by definition,
$\bar{f}_0(t)\equiv 1$) and the Callan-Treiman point $\Delta_{K\pi}$
as subtraction points leads to:
\begin{equation}
\bar {f}_0(t) =
\exp\Bigl{[}\frac{t}{\Delta_{K\pi}}(\mathrm{ln}C- G(t))\Bigr{]}~, 
\label{Dispf}
\end{equation}
with
\begin{equation} 
G(t)=\frac{\Delta_{K\pi}(\Delta_{K\pi}-t)}{\pi} \int_{t_{K \pi}}^{\infty}
\frac{ds}{s}
\frac{\phi_0(s)}
{(s-\Delta_{K\pi})(s-t-i\epsilon)}~.
\label{G}
\end{equation}
Here, 
$\phi_0(s)$ is the phase of $\bar{f}_0(s)$.  In writing
Eq.~(\ref{Dispf}), it has been assumed that $\bar{f}_0(t)$ has no zeros.
We will come back to this point  in
Sec.~\ref{Sec:zero}. In what follows, $G(t)$ is decomposed as
\begin{equation}
G(t)=G_{K \pi} (\Lambda_S,t) + G_{as} (\Lambda_S,t) \pm \delta G(t)~,
\end{equation}
where the first term corresponds to an integration from the threshold $t_{K\pi}$ 
up to a cut-off
$\Lambda_S$ which characterizes the end of the elastic region while in
the second term the integration runs from $\Lambda_S$ to $\infty$. The
choice of the value of $\Lambda_S$ will be discussed later (see
section \ref{Sec:Cutoff}).  In the elastic or low-energy region
($t_{K\pi} < s< \Lambda_S$) the phase is identified with the $s$-wave,
$I=1/2$ $K\pi$ scattering phase, $\delta_{0}$, according to Watson's
theorem \cite{Watson}.  In the analysis of Ref.~\cite{Bernard:06}, $\delta_{0}$
(with its uncertainty) has been taken from
Ref.~\cite{Buettiker:2003pp}. There a matching of the solution of the 
Roy-Steiner equations with the $K \pi \to K \pi$, $ \pi \pi \to K \bar
K$ and $\pi \pi \to \pi \pi$ scattering data available at higher
energies has been performed.  The phase obtained in this way is in
very good agreement with the work of Ref. \cite{jop00}, see below. 
In the inelastic or high-energy region ($ s > \Lambda_S$), the phase is
almost unknown. Perturbative QCD indicates \cite{Lepage} that
$\bar{f}_0(t)$ vanishes as $\calo(1/t)$ for large negative $t$.  Thus
from Eq. (\ref{G}), we conclude that the phase must approach $\pi$
asymptotically. In Ref. \cite{Bernard:06}, the phase has
been taken constant and equal to its asymptotic value of $\pi$ for $s
>\Lambda_S$ and an uncertainty of $\pm \pi$ has been assumed.  Note
that this uncertainty is a rather conservative estimate leading to a
large band going from $0$ to $2\pi$.  However, due to the two
subtractions $G(t)$ converges rapidly, hence $G(t)$ is almost
insensitive to the high energy behaviour of the phase, cf.
Sec.~\ref{Sec:Cutoff}, and the large uncertainty on the phase at high
energy turns into a small uncertainty on $G(t)$. 
For example, $G(0)$,
which has the largest error, is given for $\Lambda_S= 2.77$ GeV$^2$ by
\begin{equation} 
G(0)=0.0398 \pm 0.0018 \pm 0.0036\pm 0.0017~,
\label{Gz}
\end{equation}
where the first/second error correspond to the error on $G_{K\pi}$/$G_{as}$, 
respectively, and the third error comes form the study of
isospin breaking.  We will come back to the uncertainties on $G(t)$ in
detail in section \ref{Sec:Inputs}.

The only free parameter, $\ln C$, in Eq. (\ref{Dispf}) could, in
principle, be determined from the sum rule
\begin{equation}
\mathrm{ln}C=G(- \infty )\equiv \frac{\Delta_{K\pi}}{\pi} \ \int_{t_{\pi K}}^{\infty}
\frac{ds}{s} \frac{\phi_0(s)}{(s-\Delta_{K\pi})}~,
\label{Sumrule}
\end{equation}
dictated by the asymptotic behaviour of $\bar{f}_0(t)$,
cf. Ref.~\cite{Lepage}. However, this sum rule, which exhibits one less
subtraction than $G(t)$, Eq.~(\ref{G}), is not precise enough to allow
to determine $\ln C$ with  a good accuracy without adding any information
on the high energy behaviour of the phase of the form factor. We will
come back to the discussion of this sum rule later. 
Thus, $\ln C$ is a free parameter which can  be
determined from experiment by fitting the $K_{\mu3}$-decay
distribution with the dispersive formula for $\bar{f}_0(t)$,
Eq.~(\ref{Dispf}). 
For more details on the dispersive representation of the scalar form
factor, see Ref. \cite{Bernard:06}.

\subsection{Vector form factor}
\label{sec:vector}
In connection with the recent precise measurements of the differential 
spectrum of the  $\tau \to \nu_\tau K \pi$ decay by Belle \cite{Belle:07} and 
BaBar \cite{Babar:07}, 
theoretical work has been devoted in the last few years to the description of 
$f_+(t)$. 
Form factors have been obtained in the framework of resonance chiral theory
with additional constraints from dispersion relations in Refs. 
\cite{Jamin:2008qg,Boito:2008fq}. In Ref. \cite{Bachir08}, a coupled
channel analysis has been performed taking into account, through
analyticity requirements, the experimental information on elastic and
inelastic $K \pi$ scattering from the LASS collaboration. 
All these studies impose constraints from short distance QCD as well as the
value of the vector form factor at zero momentum transfer. A fit to
the $\tau$ data allows them to determine completely the shape of the
form factor and thus to deduce a value for the slope and the curvature
which compares reasonably well with the recent $K_{\ell 3}$
experiments. In these works, the emphasis is put on the energy
region of the $\tau$-decay  and  they are thus best suited for it. 
Here our aim is somewhat different: we want to have a very precise 
parametrization of
the form factor at low energy improving on the pole parametrization
usually assumed in the $K_{\ell 3}$-analysis
\begin{equation}
\bar{f}_{+}(t)= \frac{M_{V}^2}{M_{V}^2-t}~,
\end{equation}
which expresses the vector form factor completely in terms of a
resonance described as a discrete pole at $\sqrt{t}=M_{V}$. 
This  parametrization is
physically motivated by the dominance of the $K^*(892)$ resonance in
the vector channel. We will add to our knowledge of the presence of
this resonance, the properties of analyticity and a proper behaviour of
the phase at threshold. Contrary to the analysis discussed previously, 
we will not be able to determine the slope of the vector form factor
but it will be a free parameter to be determined from a fit
to the $K_{\ell 3}$-data. This will allow us, using a twice subtracted
dispersion relation as in the scalar case, to minimize the effect of
the high energy region in the dispersive integral over the phase of the form
factor.

In the case of the vector form factor, the value of $\bar{f}_+(t)$ at
$t=0$ is known, see Eq. (\ref{defnffactor}), but there is no
equivalence of the low-energy theorem of Callan and Treiman, Eq. (\ref{C}). 
Thus a
dispersion relation for ln$\bar{f}_+(t)$, this time twice-subtracted at
zero, will be written. Defining $|d\bar{f}_+(t)/dt|_{t=0}\equiv
\Lambda_+/m_\pi^2$ and assuming again that the form factor has no zero, this
will be discussed in section \ref{Sec:zero},  one has:
\begin{equation}
\bar{f}_+(t)=\exp\Bigl{[}\frac{t}{m_\pi^2}\left(\Lambda_+ + H(t)\right)\Bigr{]}~,~\mathrm{where}~~ 
H(t)=\frac{m_\pi^2t}{\pi} \int_{t_{K\pi}}^{\infty}
\frac{ds}{s^2}
\frac{\phi_1(s)}
{(s-t-i\epsilon)}~,
\label{Dispfp}
\end{equation}
with  $\phi_1(s)$ being the phase of $\bar{f}_+(s)$.
At sufficiently low energies, in the elastic region, 
$\phi_1(s)$ is  given} by the $p$-wave, I = 1/2 $K \pi$ scattering
phase, $\delta_{K \pi}^{l=1,I=1/2}(s) \equiv \delta_1(s)$. 
A detailed partial-wave analysis of $K \pi \to K \pi$ scattering
in the energy range $s_0 \equiv (0.825~\mathrm{GeV})^2 \leq s \leq (2.5~\mathrm{GeV})^2$
has been performed in Ref. \cite{Buettiker:2003pp} based on high statistics production experiments. 
%
In order to reliably evaluate the dispersion integral,
Eq.~(\ref{Dispfp}), an accurate extrapolation of the  scattering phase down to
threshold is needed. Contrary to the $s$-wave case,
the Roy-Steiner equations are not really useful for providing
such an extrapolation due to the lack of relevant experimental results.
However, the well known method due to Gounaris and
Sakurai~\cite{Gounaris:1968mw} can be used to directly construct a
partial wave amplitude which is unitary, has the correct threshold
behaviour, the correct analyticity properties (neglecting the left-hand
cut) and reproduces the position and width of the $K^*(892)$ as given
by the PDG~\cite{PDG:08}.
It is most suited to use the inverse amplitude method. Defining the
function $D(s)$ via the $T$-matrix as
\begin{equation}
T=\frac{q_{K\pi}^2(s)}{D(s)}~,
\end{equation}
allows to determine its discontinuity
\begin{equation}
{\rm {Im}}  \, D(s) = -2 \frac{q_{K\pi}^3}{\sqrt{s}}~.
\end{equation}
Hence, one can write a dispersive representation for $1/T$ leading to
\begin{equation}
D(s) = -2 q_{K\pi}^2(s) \frac{s}{\pi} \int_{t_{K \pi}}^\infty \frac{dx}{x} \frac{q_{K\pi}(x)}{\sqrt {x}} \frac{1}{x-s}
+P(s)
\label{eq:ps}
\end{equation}
with $P(s)$ being a subtraction polynomial.
 In the previous equations, we have used the standard notations,
\begin{equation}
s  = (p_K + p_\pi )^2, \quad q_{K\pi}(s)= \frac{\left((s-(m_K+m_\pi)^2)\, (s-(m_K-m_\pi)^2)\right)^{1/2}}{2 \sqrt{s}}~,
\label{kinvff}
\end{equation}
where  $q_{K\pi}$ is the absolute value of the three-momentum in
the $K \pi $ center of mass frame. 
Thus the $K \pi$ scattering phase is given by:
\begin{equation}
\frac{q_{K\pi}^3(s)}{\sqrt{s}} \,\mathrm{ctg}(\delta_1(s)) = q_{K\pi}^2(s) h(s) + \frac{P(s)}{2}~,
\label{eq:phase}
\end{equation}
with 
\begin{equation}
h(s) = -\frac{s}{\pi} \mathrm{P}\int_{t_{K \pi}}^\infty \frac{dx}{x} \frac{q_{K\pi}(x)}{\sqrt{x}} \frac{1}{x-s}~.
\end{equation}
A minimal choice will be done here for $P(s)$ 
namely
\begin{equation}
P(s)=a+b s~,
\label{eq:pslinear}
\end{equation}
since it already gives a very good
description of the phase in the vicinity of the resonance $K^*(892)$ and down to
threshold. We will consider the impact of a higher order polynomial
in section \ref{Sec:Vector}.  
The constants
$a$ and $b$ are determined from the mass and
the width of the  $K^*(892)$ characterized as
\begin{equation}
\mathrm{ctg}(\delta_1(s))|_{s = M_{K^*}^2} = 0\quad\mathrm{and}
\left.\quad\frac{d\delta_1(s)}{ds}\right|_{s=M_{K^*}^2} = \frac{1}{M_{K^*}\Gamma_{K^*}}~.
\label{delta1}
\end{equation}     
Note that there exists in the literature another definition of mass and width 
 in terms of the position of the pole in the complex plane. The latter is process independent.   
The uncertainties coming from 
the inputs used for $M_{K^*}$ and $\Gamma_{K^*}$ will be discussed in section
\ref{Sec:Vector}. Another possibility would be 
to determine $a$ and $b$ from a direct fit to the data \cite{Aston:1987ir}.
This would, 
however, lead to a function $H(t)$ lying within the error bars
discussed below.  
We checked that the phase constructed in this way with no free
parameters leads to values of the $p$-wave scattering length,
$a_1=0.0183 m_\pi^2$, agreeing with other determinations
\cite{Buettiker:2003pp,Bernard:1990kw,roessl99}.

\begin{figure}[h!!]
\begin{center}
\psfrag{3.14}{$\pi$}
\psfrag{6.28}{$2 \pi$}
\psfrag{9.42}{$3 \pi$}
\psfrag{0.314}{$\frac{\pi}{10}$}
\psfrag{0.628}{$\frac{2\pi}{10}$}
\psfrag{0.942}{$\frac{3\pi}{10}$}
\includegraphics*[scale=0.35,angle =0]{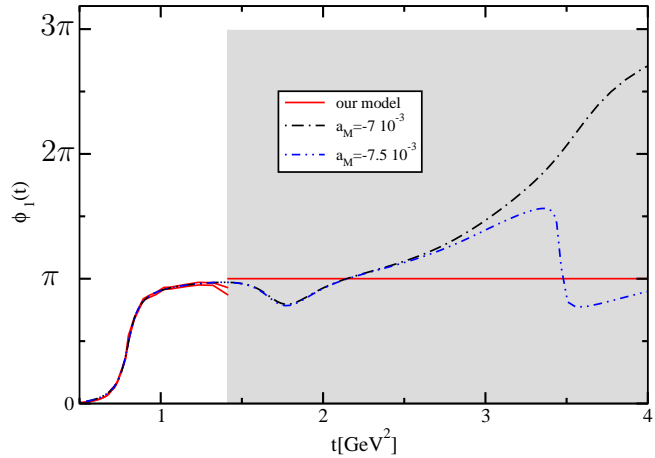}
\end{center}
\caption{\it Comparison of our model for the phase of the vector form factor, 
Eq.~(\ref{eq:phase}), with the coupled channel analysis of Ref.~\cite{Bachir08}.
{\it The grey band corresponds to the assumption that above $\Lambda_V$ 
the phase equals $\pi^{+ 2 \pi}_{- \pi}$, see text.}
\label{phasemouss}}
\end{figure}

In $p$-wave scattering, inelasticity effects which imply $\phi_1 (s)
\neq \delta_{1}(s)$ become important at lower energies than in the
scalar case, the mass of the vector resonance $K^*(1414)$ being an
indication of the start of the inelasticity.  At high energy, 
following the same arguments on 
the asymptotic behaviour as for the scalar case, the phase will
reach its asymptotic value, $\pi$. Therefore,  
similarly to what has been done for $G(t)$, the
function $H(t)$, Eq.~(\ref{Dispfp}), is decomposed into two parts:
\begin{equation}
H(t)=H_{K\pi}(\Lambda_V,t) + H_{as}(\Lambda_V,t) \pm \delta H(t)~,
\label{H decomposed}
\end{equation}
with
\begin{equation}
H_{K \pi}(\Lambda_V, t)= \frac{m_{\pi}^2~t}{\pi}\int_{t_{K \pi}}^{s_0} \frac{\delta_1(t')}{t'^2(t'-t)}dt'+\frac{m_{\pi}^2~t}{\pi}\int_{s_0}^{\Lambda_V} \frac{\delta_{exp}(t')}{t'^2(t'-t)}dt'~,
\label{dispfp}
\end{equation}
and 
\begin{equation}
H_{as}(\Lambda_V, t)=\frac{m_{\pi}^2~t}{\pi}\int^\infty_{\Lambda_V} 
\frac{\pi}{t'^2(t'-t)}dt' =  -\frac{m_{\pi}^2}{t}~\mathrm{ln}\Bigl{(}1-\frac{t}{\Lambda_V}\Bigr{)}-\frac{m_{\pi}^2}{\Lambda_V}~.
\label{Has}
\end{equation}
In these equations, $\Lambda_V$ denotes the end of the elastic region.
In  what follows, we will use $\Lambda_V=(1.414)^2~\mathrm{GeV}^2$
and we will discuss other values for $\Lambda_V$ in section \ref{Sec:Vector}. In
Eq.~(\ref{dispfp}), the analytic formula for the phase $\delta_1$,
Eq.~(\ref{eq:phase}), is used below $s_0$. Above $s_0$ and below
$\Lambda_V$ the experimental points of Aston et al.~\cite{Aston:1987ir}
are used to determine $\delta_{exp}$ and to evaluate the corresponding
contribution to $H(t)$ together with its uncertainties. As in the
scalar case, the asymptotic contribution, Eq.~(\ref{Has}), gives only a 
very small contribution to $H(t)$ due to the two subtractions. 
For the error, we will take a somewhat larger band than in the scalar case. 
Indeed it was found in Ref. \cite{Bachir08} that a fit to 
the $\tau$ data
reproducing the rate $R_{K \pi}$ quoted by the PDG \cite{PDG:08}
and compatible with asymptotic QCD, leads to a phase of
the vector form factor reaching $3 \pi$ at infinity. Thus we will take
\begin{equation}
\delta H_{as}(\Lambda_V, t) = ^{+ 2H_{as}(\Lambda_V,t)}_{- H_{as}(\Lambda_V,t)}~,
\end{equation}
corresponding to the assumption that above $\Lambda_V$, $\phi_1(s) = \pi
^{+2\pi}_{-\pi}$. This recipe almost certainly overestimates the real
uncertainty.  The other sources of uncertainty will be discussed in
section \ref{Sec:Vector}. Let us give here only as an indication 
the value of $H$ at $t_0$, point in the physical region of 
$K_{\ell3}$-decay where it has the largest error. For 
$\Lambda_V = (1.414~\mathrm{GeV})^2$, one has
\begin{equation}
H(t_0) = (2.16 \pm 0.04 ^{+0.65}_{-0.33} )\times 10^{-3}~,
\end{equation}
where the first uncertainty is the one on $H_{K\pi}$ and the second one 
the one on $H_{as}$, Eq.~(\ref{H decomposed}). 
Note that as for $G(t)$, even if the inelasticity sets in at lower
energies, the uncertainty on the value 
of $H_{as}(\Lambda_V,t)$ related to our
poor knowledge of $\phi_1(s)$ for $s>\Lambda_V$ 
is small due to the two subtractions performed this time at zero.
The resulting phase, $\phi_1(t)$, with its uncertainties (grey band) is displayed in
Fig.~\ref{phasemouss} using the charged $K^*$ mass, $M_{K^*}
=891.66$~MeV and width $\Gamma_{K^*} = 50.8$~MeV from PDG \cite{PDG:08}. 
It is compared with
the  phase obtained in Ref. \cite{Bachir08} from two  
fits characterised by a different value of the parameter
$a_{\mathrm{M}}$, see Ref. \cite{Bachir08} for more details.

Finally, let us
note that a sum rule comparable to Eq. (\ref{Sumrule}) is implied by
the asymptotic behaviour of the form factor,
\begin{equation}
\Lambda_+= -H(-\infty)= -\frac{m_\pi^2}{\pi} \int_{t_{K\pi}}^{\infty}
ds \frac{\phi_1 (s)}{s^2}~.
\end{equation}

\section{Discussion of errors and assumptions}
\label{sec:error}
The aim of the dispersive representation is to describe the shape of
the form factor in the phy-sical region of $K_{\ell3}$-decays as
precisely as possible and consequently to test the Standard Model
through the value of the scalar form factor at the Callan-Treiman
point. In order to  achieve this, it is mandatory to have under control all the
assumptions entering the construction of the form factors as well as
the determination of the errors. In the following, we will discuss the
choice of the cut-off,  the dependence on the
input parameters and the absence of zeros.

\subsection{Cut-off dependence}
\label{Sec:Cutoff}
As just discussed two energy regions are distinguished in the
dispersive analysis of the $K \pi$ form factors, the elastic one at
low energy which is very well under control and the inelastic one at
higher energies which is of less importance in the description of the
form factors due to the two subtractions in the dispersive
representation.  As explained before, in the latter region a rough
estimate of the phase is made, using $\phi_0(t)= \pi \pm \pi$ for the
scalar form factor and $\phi_1(t)=\pi ^{+2\pi}_{-\pi}$ for the vector
one.  Let us first discuss this approximation in a bit more detail.

\begin{figure}[h!]
\begin{center}
\psfrag{XXXXXXXXXXXXX}{{\scriptsize {f.f. \cite{jop00}}}}
\psfrag{XXX}{{\scriptsize{$\pi K$ amp. \cite{Buettiker:2003pp}}}}
\psfrag{XXXX}{{\scriptsize{$\pi K$ S-matrix \cite{Buettiker:2003pp}}}}
\includegraphics*[scale=0.4]{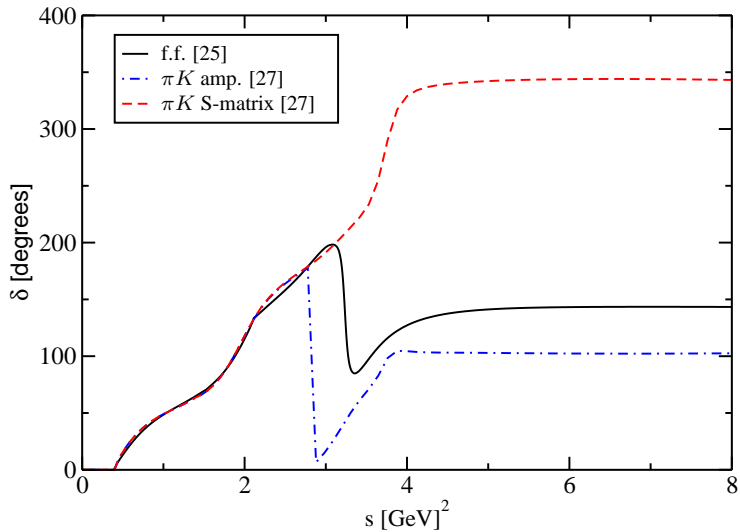}
\caption{{\it{
Comparison of the K$\pi$ scattering phases from the amplitude and 
the S-matrix in the scalar channel  extracted
in Ref.~\cite{Buettiker:2003pp} 
and the  phase of the form factor (preferred fit 6.10K2) obtained via a multi-channel analysis~\cite{jop00}.}}}
\label{fig:phase}
\end{center}
\end{figure}

In Fig.~\ref{fig:phase}, we show the phase of the scalar form factor
as obtained from a once subtracted dispersion relation by Jamin et
al., Ref. \cite{jop00}, as well as the phase of the $K \pi$ amplitude
obtained by B\"uttiker et al., Ref. \cite{Buettiker:2003pp}.
As can be seen, in the elastic region both phases agree as they should.
In the region where  the inelasticity sets in both phases decrease, even
rather abruptly in the case of the phase of the amplitude, and then
start to grow again. This behaviour is well understood and has been
explained in Refs. \cite{Bern:04, Oller:07} in the case of the scalar $\pi
\pi$ form factor. Even though the central value of the phase 
discussed in section \ref{sec:scalar} does not have this property, 
the large uncertainty assigned to it takes it into account. 
Unless the form factor has
a zero at some higher energy, as it will be discussed in the next section, 
no other sharp drop of the phase is expected and the phase will just
in some way go to its asymptotic value $\pi$ at very large $t$ as
typically does the phase obtained by Jamin et al. in Ref. \cite{jop00}. Thus
$\phi_0(t)= \pi \pm \pi$ certainly encompasses the physical phase of the
scalar form factor.

\begin{figure}[h!]
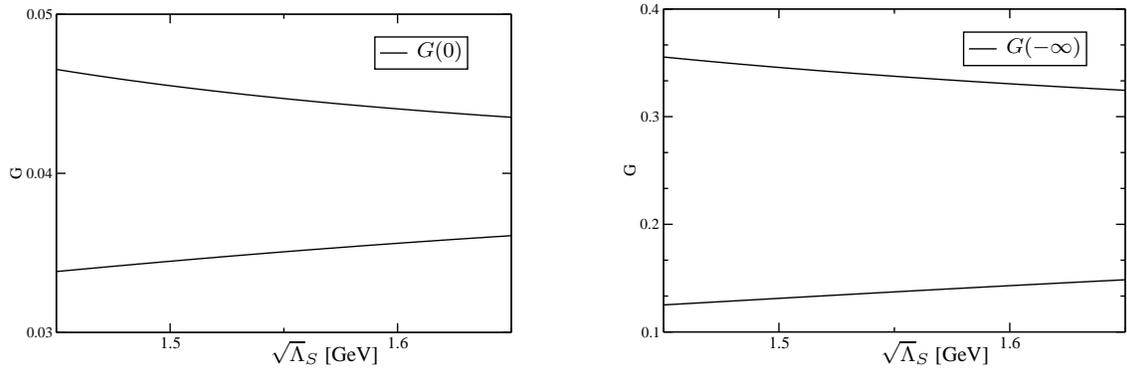

\begin{center}
\psfrag{xxx}{\scriptsize{$\sqrt \Lambda_S$ [GeV]}}
\psfrag{G ( 0  )}{\scriptsize{$G(0)$}}
\psfrag{G ( -infty  )}{\scriptsize{$G(-\infty)$}}
\includegraphics*[width=6.7cm,angle=0]{Glamb2min.eps}
\hspace{1.25cm}
\vspace{-4cm}
\includegraphics*[width=6.7cm]{Glamb2max.eps}
\vspace{4cm}
\caption{\it Variations of $G(0)$ (LHS) and $G(-\infty)$ (RHS) as a function of the 
end point of the elastic region $\Lambda_S$. Note that the  scale differs in
the two figures.} 
\label{Glamb}
\end{center} 
\end{figure}

Another source of uncertainty comes from the fact that the energies
$\Lambda_S$ and $\Lambda_V$ where the inelasticity cannot be neglected
any more are not very well known. For the $s$-wave, $\Lambda_S$ was
chosen in Ref.  \cite{Bernard:06} as the energy where the phase of the
amplitude is experimentally found to be different from the phase of
the $S$-matrix, namely at $s=(1.66~\mathrm{GeV})^2$ as shown in
Fig.~\ref{fig:phase}.  In the case of the vector form factor, the
$K^*(1414)$ resonance  can be seen as an indication of the end of
  the elastic region. We want to investigate here how sensitive
$G(t)$ and $H(t)$ are to variations of the cutoffs $\Lambda_S$ and
$\Lambda_V$ within reasonable bounds. We will concentrate on $G(t)$,
an analogous study on $H(t)$ leads to similar conclusions. In
Fig.~\ref{Glamb}, the bands represent the possible values of $G(0)$,
Eq.~(\ref{G}), and of $G(-\infty)$, Eq.~(\ref{Sumrule}), as a
function of the cut-off where $G(0)$ has the
largest uncertainty in the whole physical region of $K_{\ell3}$-decays.
For the $s$-wave a reasonable range of values for $\Lambda_S$ is ($1.43$
GeV$)^2 < \Lambda_S < (1.66$ GeV$)^2$ where the lowest value is
determined by the $K^*(1430)$ resonance.  Within these limits
$G_{min}(0)$ varies between 0.0331 and 0.0354 and $G_{max}(0)$ between
0.0474 and 0.0442 while $G_{min}(-\infty)$ varies between 0.1227 and 
0.1495 and $G_{max}(-\infty)$ between 0.3599 and 0.3234 where
$G_{max/min}(t) =G(t) \pm \delta G(t)$ and the plus sign corresponds
to the maximum value.  Hence although each part of the integral
naturally depends on the exact cut-off value, the sum is
clearly less sensitive on the exact position of $\Lambda_S$.  
We obtain, for instance,
$G(0) = 0.0336+0.0067 =0.0403$ for $\Lambda_S=(1.43$~GeV$)^2$ and $G(0)
=0.0362+0.0036 = 0.0398$ for $\Lambda_S=(1.66$ GeV$)^2$. 
As expected $G(-\infty)$, which involves one less subtraction, depends
much more on the exact value of the cut-off and has a much larger
uncertainty. However 
a comparison of the theoretical result for
$\ln C$, Eq.
(\ref{Sumrule}), together with its experimental determinations
\cite{NA48mu,KLOEmu,KTeVmu}, which lie between 0.14 to 0.21 show that the
uncertainty on $G(t)$ is certainly overestimated.  As already stated,
a precise determination of $\ln C$ from the sum rule, Eq.
(\ref{Sumrule}), is in the actual state of the art not possible.

\subsection{Variation of the input parameters}
\label{Sec:Inputs}
Within this section we  will discuss the influence of the choice of the input
parameters on the phase at low energies  and therefore on the 
dispersive representation of the form factors.

\subsubsection{Scalar form factor}
\label{Sec:Scalar}
At low energy the $I = 1/2$, $s$-wave $K\pi$ scattering phase,
$\delta_0$ from Ref.~\cite{Buettiker:2003pp} is used in the
evaluation of the function $G(t)$. As already discussed in
Ref.~\cite{Bernard:06}, there are two sources of uncertainties on
$\delta_0$. The first one comes from the propagation of the
errors  of the experimental inputs into the solution of the Roy-Steiner
equations. The other one is due to the choice of the point where one
matches the Roy-Steiner equations with the data.  This leads to the
conservative estimate (cf. Ref.~\cite{Bernard:06}),
\begin{equation}
\delta G_{K\pi}(\Lambda_S,t) \le 0.05~G_{K\pi}(\Lambda_S,t)~.
\label{dGkpi}
\end{equation}

Furthermore, the analysis of Ref.~\cite{Buettiker:2003pp} is  performed under
the assumption of perfect isospin symmetry. In
Ref.~\cite{Gasser:2007de}, it has been shown that for $\pi\pi$
scattering the shift in the phase due to isospin breaking corrections
can amount up to 0.015. For the $\pi\pi$ system,  these leading
effects are of purely electromagnetic origin. In
Refs.~\cite{Kubis:2001ij,Nehme:2001wf} it has been argued that due to a
partial cancellation between the strong and the
electromagnetic effects, the total isospin breaking effect for $K\pi$
should be smaller than for $\pi\pi$. Awaiting a more detailed
quantitative analysis~\cite{BKOP08}, we will assume here as a
conservative estimate of isospin breaking effects a constant shift of
the phase by 0.02 which is even larger than the maximal value obtained
for $\pi\pi$~\cite{Gasser:2007de}. The corresponding error on
$G_{K\pi}(\Lambda_S,t)$ is, in fact, comparable to $\delta G_{K\pi}$, 
Eq.~(\ref{dGkpi}). 
Replacing the average values of $m_K$ and
$m_\pi$ throughout the solution of Roy-Steiner equations with their
physical ones 
has only a negligible effect
on $G_{K\pi}(\Lambda_S,t)$. 

\begin{figure}[t]
\begin{center}
\includegraphics*[scale=0.35,angle=0]{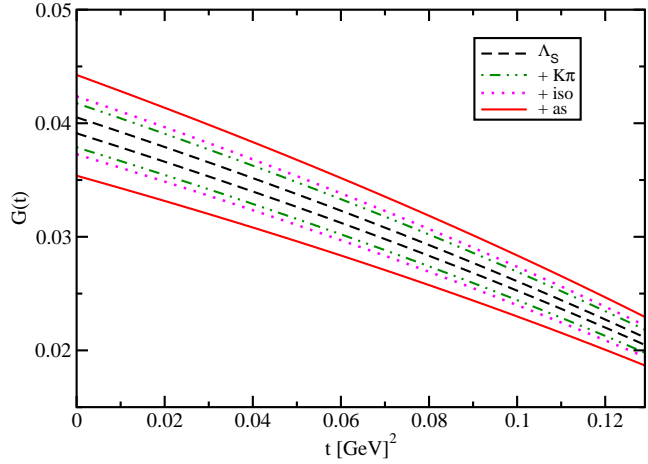}
\caption{\it The function $G(t)$ in the physical region. 
Each curve corresponds to one of the uncertainty discussed in the text
added in quadrature to the previous one, in the order shown in the 
legend. The last curve gives the total uncertainty on $G(t)$.}
\label{Gfig}
\end{center}
\end{figure}

The function $G(t)$ with the different uncertainties discussed here
is plotted in Fig.~\ref{Gfig}.
In section \ref{Sec:GP} the
simplified expression for $G(t)$
discussed in Ref. \cite{Bernard:06} 
is given together with the different errors on the parameters.

\begin{figure}[t]
\begin{center}
\psfrag{0.785}{$\frac{\pi}{4}$}
\psfrag{1.57}{$\frac{\pi}{2}$}
\psfrag{2.355}{$\frac{3\pi}{2}$}
\psfrag{3.14}{$\pi$}
\psfrag{XX}{\scriptsize{$-0.15$ GeV$^{-2}$}}
\psfrag{XXX}{\scriptsize{0}}
\psfrag{XXXXXXXX}{\scriptsize{$0.62$ GeV$^{-2}$}}
\includegraphics*[scale=0.4]{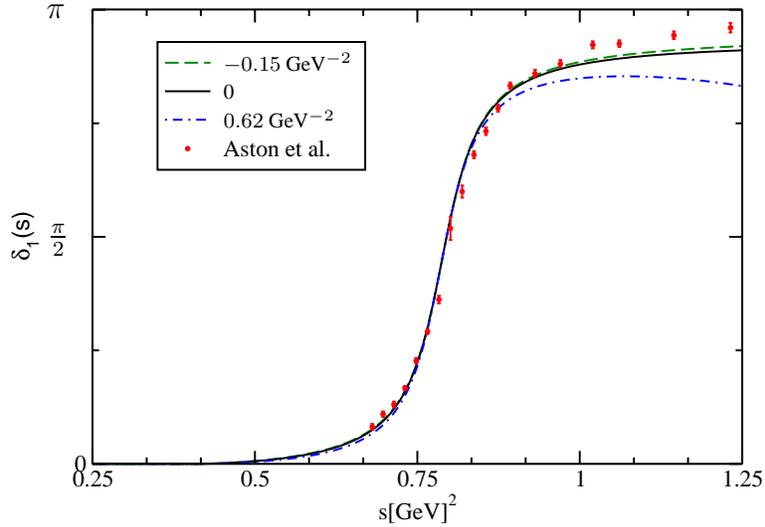}
\caption{\it 
$K \pi$ scattering phase in the vector channel for different values of
the parameter $c$. Here $s_0$ is fixed to 
the value $(0.825~\rm{GeV})^2$.  For comparison the 
data \cite{Aston:1987ir} are shown, too. 
\label{phasefp}}
\end{center}
\end{figure}

\subsubsection{Vector form factor}
\label{Sec:Vector}
To evaluate $H_{K\pi}$ we considered two domains as explained in
section~\ref{sec:vector}. From threshold $t_{K \pi}$ to $s_0$ the main
input parameters are the mass  and the width
 of the $K^*(892)$. Varying them
within
the error bars given by the PDG~\cite{PDG:08} has only a negligible
influence on $H_{K\pi}$,  the errors are at most a few 
$10^{-4}$ times $H_{K\pi}$.
Another source of uncertainty in this energy region lies in the choice
of the polynomial $P(s)$, Eq.~(\ref{eq:ps}). In order to estimate the
effect of higher order terms, we added a contribution $c \cdot s^2$ to
the r.h.s of Eq.~(\ref{eq:pslinear}) and  varied the parameter $c$
within $-0.15~{\rm GeV}^{-2} < c <0.62~{\rm GeV}^{-2}$.  
Within this range we still
obtain values for the scattering length, $a_1 \, m_\pi^3 = 0.013 -
0.020$, in agreement with other
determinations~\cite{Buettiker:2003pp,Bernard:1990kw,roessl99} and the
phases agree reasonably well with the Aston data as displayed in
Fig.~\ref{phasefp}. The induced error on $H_{K\pi}$ is in this case
a few 
$10^{-3}$ times its value. Between $s_0$ and the cutoff $\Lambda_V$ the uncertainty on
the phase is given by the error bars of the Aston data. To determine
the error induced by the choice of the cutoff values, we impose a
rather large variation for $s_0$ and $\Lambda_V$, namely
$(0.825~\mathrm{GeV})^2 < s_0 < (1.1~\mathrm{GeV})^2 $ and
$(1.2~\mathrm{GeV})^2<\Lambda_V<(1.6~\mathrm{GeV})^2$.

\begin{figure}[t]
\begin{center}
\includegraphics*[scale=0.35,angle=0]{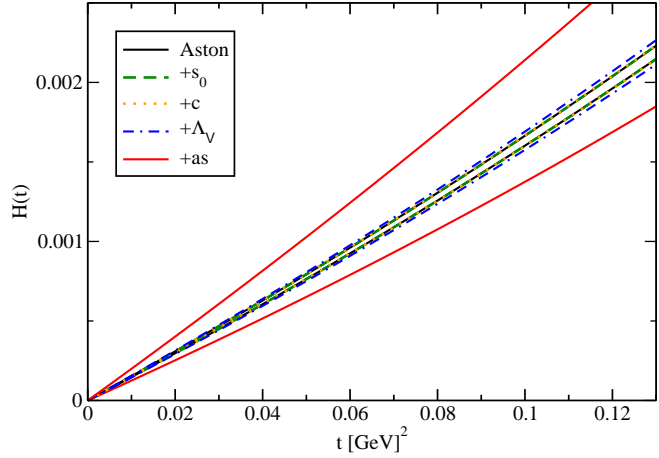}
\caption{\it The function $H(t)$. Each curve corresponds to one of the 
uncertainty discussed in the text
added in quadrature to the previous one, in the order shown in the 
legend. The last curve gives the total uncertainty on $H(t)$. }
\label{H}
\end{center}
\end{figure}
The function $H(t)$ with the different uncertainties
discussed here is plotted in Fig.~\ref{H}. 
Except for $H_{as}$, we have symmetrized the errors  which were 
not 
symmetric around the central value, using as error the largest value 
to be on the conservative side. In section \ref{Sec:HP} a
simplified expression for $H(t)$ is given together with the different
sources of uncertainties on the parameters.

\subsection {Discussion of possible zeros }
\label{Sec:zero}
Writing the dispersive relations, Eqs.~(\ref{Dispf}, \ref{Dispfp}), we
have assumed that the form factors have no zero. This is in fact quite
standard when studying form factors. Indeed, in the space-like
region, a form factor represents the Fourier transform of a charge
density. It is argued for instance in Ref. \cite{leut02} for the case of
the electromagnetic form factor of the pion that the properties of the
pion charge distribution should be similar to the one of the electron
in the ground state of the hydrogen atom. This charge density being
proportional to the square of the wave function, the corresponding form
factor is positive in the space-like region.  The possibility that the
form factors have one or several zeros can, however, not be completely
overruled and has been studied in the literature for example for the
pion form factor in Ref. \cite{rasz76}. Let us therefore discuss it here for the
scalar form factor. In the vector case the experimental situation  
is much better as we have seen in the introduction. Slope and curvature can
be measured rather precisely. Zeros would essentially modify the relation
between the slope and the curvature in the physical region of $K_{\ell3}$-decays 
compared to what we have been discussing here. Considering the  
precision of the experiment, allowed zeros cannot alter much the results
in the physical region. 
We will however briefly mention what happens for the vector 
form factor in Sec. \ref{Sec:Tau} in relation with the $\tau \to K \pi \nu_{\tau}$ decay. 
\subsubsection{Real zeros}
We will consider first one real zero, $T_0$, to simplify. 
In that case the function $\bar{f}_0(t)/(1-t/T_0)$
has no zero and one can proceed as in the 
previous section. One can thus write
\begin{equation}
\bar{f}_0(t)=  \left( 1-\frac{t}{T_0} \right)\exp\left[\frac{t}{\Delta_{K\pi}}\left(\mathrm{ln}C- 
\ln\left(1-\frac{\Delta_{K\pi}}{T_0}\right) -G(t)\right)\right]~.
\label{zero}
\end{equation}
In order for $\bar{f}_0 (t)$ to behave like $1/t$ for large negative $t$
\cite{Lepage}, the phase of the form factor has now to go to $2 \pi$ 
at infinity since one has
\begin{equation}
\lim_{t \to -\infty}\bar{f}_0(t) = \mathrm{const.}~t/t^{\delta(+ \infty)/\pi}~.
\label{phias}
\end{equation}
Now there are two possibilities. The zero can be in the time-like or in
the space-like region. 

\begin{figure}[h!!]
\begin{center}
\psfrag{yyyyyyyyyyyyyyyyyyyyyy}{\scriptsize{none}}
\psfrag{yy}{\scriptsize{$T_0=2.5$ GeV$^2$}}
\psfrag{XXX}{\scriptsize{$Z_0=(2.5 \pm {\rm i}\, 10)$ GeV$^2$}}
\psfrag{XX}{\scriptsize{$Z_0=(0.1 \pm {\rm i}\, 2)$ GeV$^2$}}
\includegraphics*[width=9.cm,angle=0]{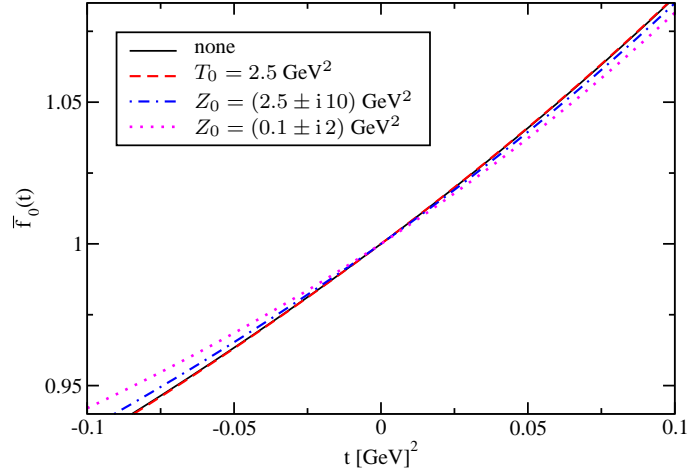}
\caption{\it Scalar form factor: without zeros plain black curve, 
with a zero at $T_0=2.5~{\rm GeV}^2$, red dashed curve and with two
complex conjugate zeros at $Z_0=(2.5 \pm {\rm i}~10 )~{\rm GeV}^2$, blue
dashed-dotted curve and at $Z_0=(0.1 \pm {\rm i}~2)~{\rm GeV}^2$ pink
dotted curve.}
\label{zerophy}
\end{center}
\end{figure}

\begin{figure}[h!!]
\begin{center}
\includegraphics*[width=9.5cm,angle=0]{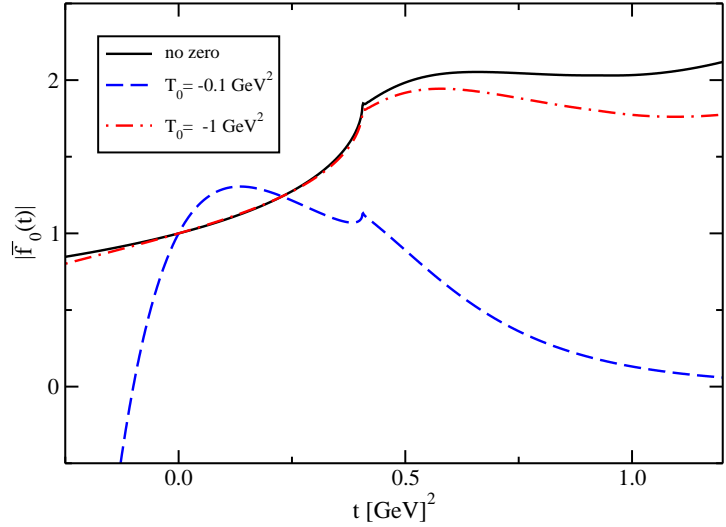}
\caption{\it Scalar form factor 
in presence of zeros located 
in the space-like region. Black plain curve: no zero,
blue dashed/red dotted-dashed curves: real negative zero at $T_0=-0.1~{\rm GeV}^2$ and 
$-1~{\rm GeV}^2$, respectively.} 
\label{zerounphy}
\end{center}
\end{figure}

\paragraph{$\bullet$ Time-like region} 
In the time-like region a zero on the real axis would correspond to 
a jump of the phase by $\pi$. In fact, Eq.~(\ref{zero}) can be rewritten as
\begin{equation}
\bar{f}_0(t)=\exp\left[ \frac{t}{\Delta_{K\pi}}\left(\mathrm{ln}C 
-\frac{\Delta_{K\pi}(\Delta_{K\pi}-t)}{\pi} \int_{t_{\pi K}}^{\infty}
\frac{ds}{s}
\frac{\phi_0(s)-\pi~\theta(s-T_0)}
{(s-\Delta_{K\pi})(s-t-i\epsilon )} \right) \right ]~.
\end{equation}
Following the discussion of section~\ref{Sec:Cutoff}, this zero should
occur at a four momentum larger than $\sim (1.7$ GeV)$^2$ since 
below, the phase of $\bar{f}_0(t)$ is known and there is no 
indication of a zero for the form factor. At these energies there is a
right-hand cut on the real axis and this scenario looks thus rather
improbable.  Furthermore since the region above $\Lambda_S$ gives a
small contribution to $G(t)$ in the $K_{\mu 3}$-physical region, one does not
expect much change in the form factor from the presence of such a
zero.  This is illustrated in Fig.~\ref{zerophy} where a comparison is made
between the form factor
obtained from Eq.~(\ref{Dispf}) and the one from Eq.~(\ref{zero}) for $T_0=2.5$
GeV$^2$. This is typically the smallest possible value for
a real zero  in this time-like region and thus the one which is expected 
to affect most our representation.
\paragraph{$\bullet$ Space-like region} 
In the space-like region, there is no left-hand cut and thus a real
zero looks more plausible even though it goes against the argument of
positivity of the form factor in the space-like region given
previously.  A completely different behaviour of the form factor as
compared with the case without zero could result if the zero was close
enough to the physical region. This can be seen in
Fig.~\ref{zerounphy} where the scalar form factor is shown for two
different values of $T_0$, namely $-0.1$ GeV$^2$ and $-1$ GeV$^2$.
In the first case, the slope is rather large and there is a maximum at
a rather small $t$ value. As $T_0$ decreases, the slope becomes
smaller and the maximum moves towards larger $t$ and eventually
approaches the curve without zero. This is in fact what happens for
$T_0=-1$ GeV$^2$.   The $K_{\mu 3}$-data, however, seem to exclude 
the large slopes we observe when the zero becomes close to the 
physical region, see Fig.~\ref{zerounphy}. 
In that case, the high energy behaviour of the form factor is completely different, too. 
This has an impact for instance in $\tau$-decays as discussed below.
\subsubsection{Complex zeros}
Zeros could occur in the complex plane, which would in fact mean, due
to the property of real-analyticity of the form-factor, the presence
of a zero, $Z_0$, and its complex conjugate $\bar{Z}_0$. The form-factor would
take the following form
\begin{equation}
\bar{f}_0(t)=  \left( 1-\frac{t}{Z_0} \right) \left( 1-\frac{t}{\bar{Z}_0} \right)
\exp\left[\frac{t}{\Delta_{K\pi}}\left(\mathrm{ln}C- 
\ln\left(1-\frac{\Delta_{K\pi}}{Z_0}\right) -\ln\left(1-\frac{\Delta_{K\pi}}{\bar{Z}_0}\right)-
G(t)\right)\right]~.
\label{zeroC}
\end{equation}
In order for $\bar{f}_0(t)$ to behave as $1/t$ for large negative $t$
\cite{Lepage}, the phase should now go asymptotically to $3 \pi$, see
Eq.~(\ref{phias}).  As can be seen in Fig.~\ref{zerophy}, the 
presence of complex zeros could lead to lower values  for the form factor at 
the end of the physical region. In the case of the blue dot-dashed curve
which corresponds to zeros with rather large imaginary parts, namely
$\pm 10$ GeV$^2$ and a rather  large real part, too, thus not
close to the physical region of  $K_{\mu 3}$-decay, the difference 
to the curve without zero is very small as expected. 
The effect becomes, however, much stronger
when one takes a small real part as well as a smaller imaginary part,
as shown by the pink dotted curve in Fig.~\ref{zerophy} which
corresponds to zeros at $(0.1 \pm \rm{i}~2)$ GeV$^2$. The slope becomes
smaller and the curvature larger. Note that the same value for $\ln C$
has been used in the determination of the curves, namely  $\ln C = 0.2138$, 
Eq.~(\ref{lnC}) 
(the trends discussed here are independent of the precise
value of $\ln C$) such that the three curves meet at the CT point.

\subsubsection{Properties of the scalar form factor and their 
constraints on the existence of possible zeros}
\label{sec:sumrules}
It is clear from the previous discussion that the presence of zeros in
the scalar form factor with values close to the physical region would
affect the determination of $\ln C$ from experiment. Unfortunately, it
is not possible to test such a scenario with the $K_{\mu 3}$-data
since due to strong correlations  and the sensitivity of the data 
it is only possible to determine one
parameter  for the scalar form factor from the fits. However, the
scalar form factor possesses some properties which might allow to set
up constraints on the presence or the absence of zeros. The different
behaviour for $t$ away from the physical region of $K_{\ell 3}$-decays
could have consequences, too, for example in $\tau$-decays.

\paragraph{$\bullet$ Sum rules}
The first properties which can be used are sum rules which can be
derived from the knowledge of the behaviour of the form factor for $t
\to -\infty$. In the absence of zeros, one obtains the sum-rule, Eq.
(\ref{Sumrule}).  It is modified in the presence of zeros and for
instance, with one real zero, it reads
\begin{equation} 
G(-\infty) = \ln C - \ln \left(1 - \frac{\Delta_{K \pi}}{T_0}\right)~.
\end{equation}
The second logarithm on the right-hand side is negative so that the
sum rule now leads to a value $G(-\infty)$ larger than $\ln C$.  For
example, with $T_0=2.5$ GeV$^2$ as used above, one obtains $\ln
(1-\Delta_{K \pi} /T_0)=-0.096$.  Thus a good knowledge of these two
quantities could provide an information on the presence of zeros in
the form factor.  In the case of complex zeros, one can derive two
additional sum-rules namely
\begin{eqnarray} 
\label{SumRule2}
\int_{t_{K \pi}}^{\infty} ds \frac{\text{Im} \bar{f}_0(s)}{\left( 1- s / Z_0 \right) \left( 1- s / \bar{Z}_0 \right)} =0~, \\
\int_{t_{K\pi}}^{\infty} ds\frac{s~\text{Im} \bar{f}_0(s)}{(1-s/Z_0)(1- s/ \bar{Z}_0)} =0~.
\label{SumRule3}
\end{eqnarray}
All these sum rules have to be satisfied simultaneously.
Unfortunately, they are rather sensitive to the phase at high energies
which, as has already been stressed, is badly known. Consequently,
they do not provide any constraint on the presence or absence of zeros
as long as one does not have a better knowledge of the behaviour of
this high energy phase. At present, it is possible to find a model for
the phase which satisfies simultaneously the three sum rules.

\paragraph {$\bullet$ Soft kaon analog of the Callan Treiman theorem} 

Another important property which provides us with a more severe constraint is
the soft kaon analog of the Callan Treiman theorem~\cite{oehme66}.
One has 
\begin{equation}
f_0(m_\pi^2-m_K^2)= \frac{F_{\pi^+}}{F_{K^+}} + \tilde \Delta_{CT}~.
\end{equation}
A one loop calculation of the $SU(3)$ correction $\tilde \Delta_{CT}$
in the isospin limit \cite{GL84} gives $\tilde \Delta_{CT}=0.03$
which is larger than its soft pion analog $\Delta_{CT}$, see
Eq.~(\ref{Delta_CT}), by a factor $m_K^2/m_\pi^2$.  It is rather small
for a first order $SU(3) \otimes SU(3)$ breaking effect, which is
expected to be of the order of $25\%$. Note one
interesting point: at NLO within the minimal not-quite decoupling
electroweak low-energy effective theory (LEET) \cite{hs04}, 
there appear in the light quark sector essentially
two combinations of parameters of spurionic origin describing the
couplings of quarks to the $W$-boson to be determined from
experiment~\cite{stern06,BOPS08}. While the knowledge of the scalar
form factor at the CT point measures one combination,  its knowledge at
$m_\pi^2-m_K^2$ measures the other one. A precise determination of
$\tilde \Delta_{CT}$ would thus help to settle the issue of the
presence of right-handed couplings of quarks to the $W$-boson.  At
present one can only give an estimate of the higher order contribution
to $\tilde \Delta_{CT}$.   The expected size is of the order of 
ten percent. To get an idea of the actual size, let us look at the 
two-loop calculation~\cite{bij}, where
one finds:
\begin{equation}
\tilde \Delta_{CT} = 2 - \frac{F_K}{F_\pi} - \frac{F_\pi}{F_K} -\frac{16}{F_\pi^4} (2 C_{12}+C_{34}) m_K^2
(m_K^2-m_\pi^2) + \bar \Delta(m_\pi^2-m_K^2) + \Delta(0)~.
\label{tilde2loop}
\end{equation}
The quantities $\bar\Delta(t)$ {\footnote {Note that in Ref.~\cite{bij} 
the parametrization of
$\bar\Delta(t)$ is in principle only valid in the physical region. 
We will however use it here in order to get an estimate for $\tilde \Delta_{CT}$.}} 
and $\Delta(0)$ are discussed in Ref.~\cite{bij}, and $C_{12}$ and $C_{ 34}$ 
are two low energy constants (LECs). 
The same combination of these LECs in Eq. (\ref{tilde2loop})
appears in the two loop calculation of the slope, $\lambda_0$, of the
scalar form factor.  Taking $\lambda_0$ between 0.009 and 0.016 which
encompasses the values obtained by the NA48, KTeV and KLOE collaboration
and using the value 1.22 for $F_K/F_\pi$ as in Ref. \cite{bij} (for an
actual status on the value of  $F_K/F_\pi$, see for example Ref. \cite{flavianet})
one obtains $-4 \times 10^{-6} < 2~C_{12} + C_{34} < 8 \times 10^{-6}$ in
agreement with the estimates found in the literature \cite{jop04, ceekpp05, KN08, BP08}. This
leads to $-0.035 < \tilde \Delta_{CT} < 0.11$ within the expected
order of magnitude. Assuming the Standard Model electroweak couplings, one has
$F_\pi / (F_K f_+(0)) =0.8752 \pm 0.0020$, such that a conservative result for the
normalized scalar form factor $\bar f_0(m_\pi^2-m_K^2)$ is
\begin{equation}
0.8< \bar f_0(m_\pi^2-m_K^2) <1~.
\label{bounds}
\end{equation}
This bound constrains the allowed region of possible zeros in the
scalar form factor. Real zeros in the space-like region which do have
an impact on the form factor in the physical region are excluded since
they drop very fast with $t$ for $t$ negative and violate the
bound.  For complex zeros the situation is less obvious.
It is easy to calculate which are the imaginary parts the complex
zeros should have in order that the scalar form factor at
$t=m_\pi^2-m_K^2$ lies within the bound, Eq.~(\ref{bounds}). One finds that the
positive imaginary part increases when the bound decreases. Typically
for a complex zero with a real part between -0.2 and 0.2 the imaginary
part varies between 0.7 for $\bar f_0(m_\pi^2-m_K^2)=1.03$ and 6 for $ \bar
f_0(m_\pi^2-m_K^2)=0.87$. Thus, with our present knowledge of $\tilde
\Delta_{CT}$, one cannot totally eliminate the presence of complex
zeros very close to the physical region which would affect the
determination of $\ln C$.

\subsubsection{High energy behaviour and the decay $\tau \rightarrow K \pi \nu_{\tau}$}
\label{Sec:Tau}
\begin{figure}[h!]
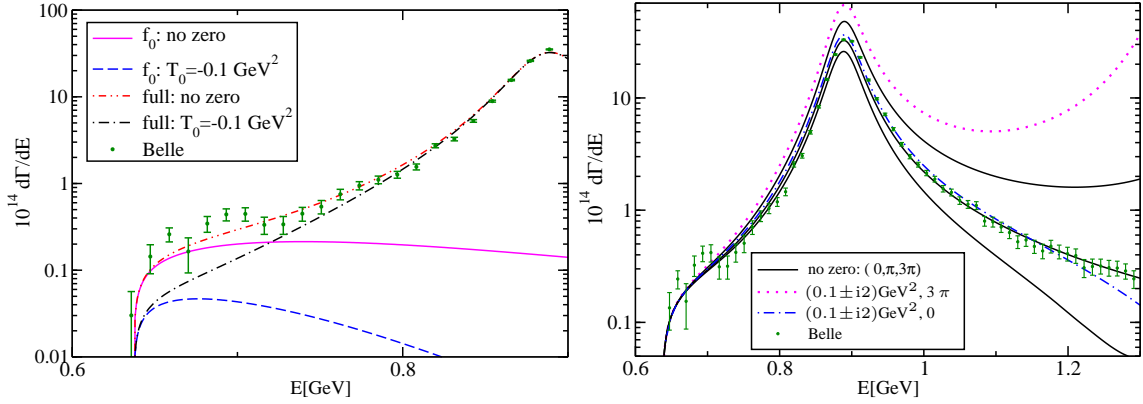

\vskip 0.45cm
\epsfysize=9.0cm
\psfrag{X}{\tiny{no zero}}
\psfrag{XXXXXXXXXXXXXXX}{\tiny{$\!\!T_0\!=\!-0.5$GeV$^2$}}
\psfrag{XXXXXXXX}{{\tiny{$\!(0.1 \! \pm \! {\rm i} 2)$GeV$^2$,  3 $\pi$ }}}
\psfrag{XXXXXXXXXXXXX}{{\tiny{$\!(0.1 \! \pm \! {\rm i} 2)$GeV$^2$,
  0}}}
\includegraphics[width=7.5cm,angle=0]{GammaTauS1.eps}
\includegraphics*[width=7.5cm,angle=0]{GammaTauvf.eps}
\caption{\it Differential decay width 
$\tau \rightarrow K \pi \nu_\tau$. On the left-hand side the calculations 
with one real zero in the scalar form factor close to the physical region 
($T_0 = -0.1~{\rm GeV}^2$) and without zero are compared. The vector 
form factor has no zero.  On the right-hand side the calculation is done 
without zero and with complex zeros at $Z_0=(0.1 \pm \rm{i} 2)~\rm{GeV}^2$ 
in the vector form factor. Different cases have been considered
corresponding to  different ans\"atze for the phase 
as explained in the text.
The scalar form factor has no zero. In these figures $\ln C = 0.2138$, 
Eq. (\ref{lnC}) and $\Lambda_+ =0.02450$ as obtained
from the pole parametrization with the K*(892) 
mass. For comparison the Belle data for 
$\tau ^- \rightarrow K_S^0 \pi^- \nu_\tau$ 
\cite{Belle:07} are displayed, too.}
\label{taukpi}
\end{figure}

The energy distribution in the decay $\tau \rightarrow K \pi \nu_\tau$ has
been measured first by the ALEPH colla-boration \cite{Barate:99}, then
by OPAL \cite{OPAL2}, and recently the Belle collaboration
\cite{Belle:07} and the BaBar one \cite{Babar:07} presented results
for different isospin combinations. The cross section is described by
\begin{eqnarray}\label{tauspeckpi}
&&{d\Gamma_{K\pi}(t)\over d\sqrt{t}} =
\frac{V_{us}^2 G_F^2 m_\tau^3 } {128 \pi^3} {q_{K\pi}(t) }
\left( 1- {t\over \mtaud} \right)^2 \times\\
&& \left[ 
\left( 1+ {2t\over \mtaud}\right) 
{4q^2_{K\pi}(t)\over t}     \vert f_+(t)\vert^2\right.
\left.
+{3(\mkd-\mpid)^2\over t^2} \vert f_0(t)\vert^2
\right]\ ,\nonumber
\end{eqnarray}
with the kinematic variable $q_{K\pi}$ defined in Eq.~(\ref{kinvff}). 
It thus involves the scalar and the vector form factors. Unfortunately
due to the presence of the $K^*(892)$ resonance which dominates the
cross section between $\sim 0.8$ and $\sim 1.2$ GeV, one can only hope to
obtain information on the scalar form factor from this decay channel
very close to threshold.
As discussed above, the only zeros which could lead to a clear difference 
in the energy distribution very close to the threshold are the real space-like zeros 
close to the $K_{\mu 3}$-physical region. 
As is shown in Fig.~\ref{taukpi} the curve is shifted towards larger values
if there is no zero in the
scalar form factor.  
The Belle data, seem to favor the latter even
though the error bars are rather large for the lowest point{\footnote{
We do not show the BaBar data since they are not publicly 
available. The same trends hold, however, for them, too.}}. This is
consistent with the soft kaon analog of the CT theorem which, as we
have seen, excludes the presence of such zeros in the scalar form
factor.

On the right-hand side of the figure, the differential decay width,
Eq. (\ref{tauspeckpi}), is shown in a broader energy range and assuming
the presence or absence of zeros in the vector form factor. The
resonance region, where the contribution of the vector form factor
dominates, becomes more sensitive to the phase in the inelastic region
such that relatively large uncertainties for the decay width are
expected in this resonance region. The dispersive representation
without zero, for $\Lambda_+ = 0.02450$, and $\phi_1 = \pi$ beyond the
cut off $\Lambda_V$, as discussed in section \ref{sec:vector},
describes amazingly well the precise Belle data in this region, see
the central plain black curve on the RHS of Fig.~\ref{taukpi}.  Much
larger or smaller phases at the beginning of the inelastic region seem
to be excluded as shown on the same figure where instead of $\pi$, $0$
for the lower plain black curve and $ 3\pi$ for the upper plain black
curve have been used for $\phi_1$. In the presence of complex zeros,
on the contrary, the phase should be very small at the beginning of
the inelastic region in order to reproduce the data, as illustrated in
Fig.~\ref{taukpi}.  There the result with a phase equal to 3 $\pi$ in
the energy range from 1.4 to 3 GeV (pink dotted curve) is compared
with the one with a vanishing phase in that same energy range (blue
dashed curve). The opposite is true for real space-like zeros.
As the value of $T_0$ increases, the resonance peak gets more and more washed
out and eventually disappears for a given phase $\phi_1$. Rather highly 
improbable large values of the phase in the inelastic region becomes 
necessary to counterbalance
the effect of the zeros. 
Thus from our study, due to the lack of knowledge of $\phi_1$ in the inelastic
(high-energy) region, we cannot
completely rule out the presence of zeros in the vector form factor
even though such a scenario does not seem very probable. 
We can, however, conclude as expected that zeros in the vector form factor 
which  
would not be totally excluded from the analysis of tau decays 
would not affect the low energy region of the vector form factor and consequently 
the results of the analysis of
$K_{\ell 3}$-decays. 
\section{Final remarks and Conclusion}
\label{sec:summary}

\begin{figure}[t]
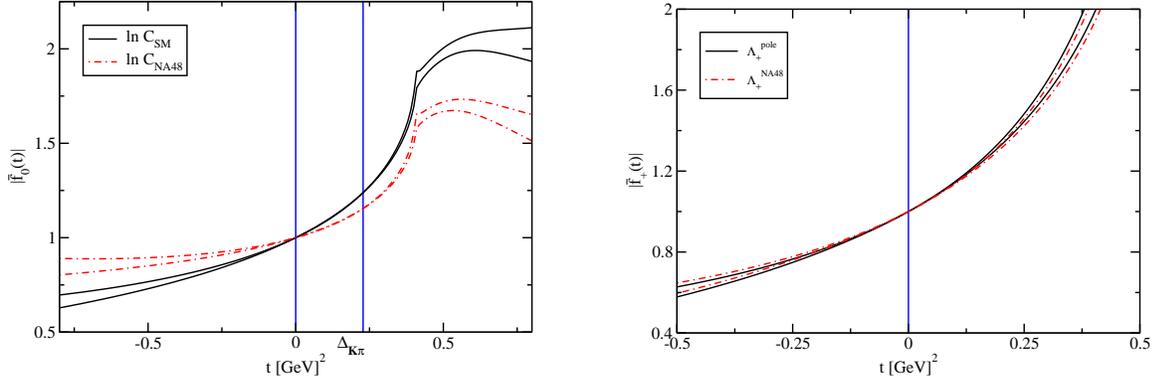

\vspace{1cm}
\includegraphics*[width=7cm]{fs.eps}
\hspace{1cm}
\includegraphics*[width=7cm]{fv.eps}
\caption{\it{Dispersive representation of the scalar  (left panel) and vector
 (right panel) form factor over
a wide range of energies. For comparison 
 two different values of  $\ln C$ (right) and 
$\Lambda_+$ (left) have been used: $\ln C|_{SM}$ is from Eq. (\ref{lnC}),
$\ln C|_{\rm{NA48}}=0.1438$,  $\Lambda_+^{\rm{NA48}}=0.0233$ are 
 the central values of the NA48 experimental results \cite{NA48mu} and 
$\Lambda_+^{pole}=0.02450$  is from the pole parametrization with the $K^*(892)$ mass. 
The band takes care of all the uncertainties discussed in the text. 
}}
\label{ffscalar}
\end{figure}

In this paper we have discussed the robustness of a precise and
convenient dispersive representation of the scalar and vector $K\pi$ form
factors.  In Fig.\ref{ffscalar} we show the scalar (left panel),
Eq. (\ref{Dispf}), and the vector form factors (right panel),
Eq. (\ref{Dispfp}), with all the uncertainties discussed 
above under the usual assumption of no zeros in the form factors and for 
two different values of their respective free parameters 
ln$C$ and $\Lambda_+$. 
The figure nicely illustrates the fact that
our dispersive representation 
describes to a very high accuracy the form factor shapes in the
physical region of $K_{\ell 3}$-decays. Eqs. (\ref{Dispf},
\ref{Dispfp}) thus represent a very useful tool for an optimal
analysis of the $K_{\ell 3}$-data. As already pointed out, it allows to
determine the shape of these form factors in an unambiguous way,
contrary to other parametrizations used in the data analysis. Furthermore, 
as emphasized in Ref. \cite{Bernard:06}, a measurement of
$\ln C$, with $C$ the value of the normalized scalar form factor at
the Callan-Treiman point, allows to test the Standard Model. A
departure of the measured value from Eq.~(\ref{lnC}) would signal, under the
hypothesis of no zeros in the form factor, a failure of the SM, as for
example the presence of a direct coupling of right-handed quarks to W
\cite{Bernard:06}.  We have, however, to moderate slightly the 
conclusion drawn there.  We have indeed seen that the shape of the
scalar form factor could be slightly modified in the highly improbable
case where it would have zeros in a very small domain of the complex
plane within or close to the $K_{\mu 3}$ physical region. Even though
the likelihood of this scenario is very small we have not been able at
present to totally eliminate it.  Note, however, that for the vector form
factor, zeros that would affect the dispersive parametrization in the 
low energy region we are interested in here, are excluded by the tau data.

\vspace{1cm}
\noindent{\bf {\large Acknowledgments:}}

We would like to thank H. Leutwyler for very useful and interesting
discussions and a careful reading of the manuscript. We are grateful to 
B. Moussallam for enjoyable discussions and to him and M. Jamin
for providing us with some of their results.  One of us (E. P.) 
would like to thank I. Caprini and P. Minkowski for very interesting discussions. 
This work has been partially supported by the EU contract MRTN-CT-2006-035482
(``Flavianet''), by the European Community-Research
Infrastructure Integrating Activity Study of Strongly Interacting 
Matter (acronym HadronPhysics2, Grant Agreement n. 227431)
under the Seventh Framework Programme of EU, by the Swiss National Science Foundation 
and by the IN2P3 theoretical project "Signature exp\'erimentale des couplages \'electrofaibles 
non-standards des quarks".
  
\begin{appendix}
\section{Some useful expressions}
\label{sec:useful}
\subsection{Scalar form factor}
\label{Sec:GP}
In order to facilitate and accelerate the numerical evaluation of the
scalar form factor for exam-ple in the experimental analysis of 
$K_{\mu3}^L$-decays, it is convenient to have a parametrization of the
function $G(t)$.  As already discussed in Ref.~\cite{Bernard:06}, within the
physical region it can be very accurately parametrized as
\begin{equation} 
G_P(t) = x D + (1-x) d + x (1-x) k~,
\end{equation}
where $x \equiv t/t_0, d \equiv G(0), D \equiv G(t_0) $. The value of
$k$ can be obtained from the constraint $G(\Delta_{K\pi}) =0$. In
Table~\ref{table:ddk} we give the central values for the parameter
$d,D,k$ together with the corresponding errors arising from the
different sources of errors as discussed in the previous section. 
\begin{table}[h!]
\begin{center}
\begin{tabular}{|l||c|c|c|c|c|c|}
\hline
 & Central value &$\delta G_{\Lambda}$& $\delta G_{as}$ &
 $\delta G_{K\pi}$ & $\delta G_{\mathit{isospin}}$& total error \\ 
\hline 
$d$& 0.0398 &0.0005 & 0.0036 & 0.0018 & 0.0017& 0.0044\\ 
\hline
$D$ &0.0209 &0.0002& 0.0016 & 0.0010&0.0010& 0.0021\\ 
\hline
$k$ &0.0045 & 0.0000& 0.0001 &0.0002&0.0003&0.0004\\ 
\hline
\end{tabular}
\end{center}
\caption {\label{table:ddk}
Coefficients arising in the parametrization $G_P$ with their uncertainties.}
\end{table}

For practical purposes and for comparison with the traditionally
often used linear and quadratic approximations to the form factor, it
is useful to list the first coefficients of the Taylor expansion of
the form factor. They read
\begin{eqnarray}
\frac{1}{m_\pi^2} \left.\frac{d\bar{f}_0}{dt}\right|_{t = 0} &=&
\frac{m_\pi^2}{\Delta_{K\pi} } (\ln C - G(0)) = 
\frac{m_\pi^2}{\Delta_{K\pi} } (\ln C - 0.0398(44))~, \nonumber \\
 \frac{1}{m_\pi^4} \left.\frac{d^2\bar{f}_0}{dt^2}\right|_{t = 0} &=&
\left(m_\pi^2 \left.\frac{d\bar{f}_0}{dt}\right|_{t =
    0}\right)^2 - 2 \frac{m_\pi^4}{\Delta_{K\pi}} G'(0) = 
\left(m_\pi^2 \left.\frac{d\bar{f}_0}{dt}\right|_{t =
    0}\right)^2 + (4.16\pm 0.56)\times 10^{-4}~, \nonumber \\
 \frac{1}{m_\pi^6} \left.\frac{d^3\bar{f}_0}{dt^3}\right|_{t = 0} &=&
\left(m_\pi^2 \left.\frac{d\bar{f}_0}{dt}\right|_{t =
    0}\right)^3 - 6 \frac{m_\pi^4}{\Delta_{K\pi}} G'(0) 
\left(m_\pi^2 \left.\frac{d\bar{f}_0}{dt}\right|_{t =
    0}\right) - 3 \frac{m_\pi^6}{\Delta_{K\pi}} G''(0) \nonumber \\ &=& 
\left(m_\pi^2 \left.\frac{d\bar{f}_0}{dt}\right|_{t =
    0}\right)^3 + 3\, (4.16\pm 0.56)\times 10^{-4}\, 
\left(m_\pi^2 \left.\frac{d\bar{f}_0}{dt}\right|_{t =
    0}\right) + (2.72 \pm 0.21)\times 10^{-5}~. \nonumber \\
\end{eqnarray} 
The Taylor expansion up to third order of the scalar form factor
allows to well reproduce the exact dispersive representation in
the physical region. The maximal error is $3\%$. 
\subsection{Vector form factor}
\label{Sec:HP}
It is equally very convenient to have a parametrisation of the
function $H(t)$ in order to avoid the evaluation of the dispersive
integral for the vector form factor. Within the physical region, the
function $H(t)$ can be very accurately parametrized as
\begin{equation}
H_P(t) = H_1 x + H_2 x^2~.
\end{equation}
The numerical values of the parameters $H_1$ and $H_2$ are given in
Table~\ref{table:h1h2} together with the corres-ponding uncertainties. As discussed 
in the text, the uncertainties coming from the uncertainties on the mass and the width of the $K^*$, 
$\delta H_{M_{K^*}}$ and $\delta H_{\Gamma_{K^*}}$ are completely negligible.  
\begin{table}[h!]
\begin{center}
\begin{tabular}{|l||c|c|c|c|c|c|c|}
\hline
 & Central value &  $\delta H_{Aston}$ & $\delta H_{s_0}$ & $\delta H_c$ & 
$\delta H_{\Lambda_V}$& $\delta H_{as}$ &
total error \\ 
\hline 
$H_1\times 10^3$& 1.92 & 0.04 & 0.01 & 0.01 & 0.06 & $^{+0.62}_{-0.31}$ & $^{+0.63}_{-0.32}$ \\ 
\hline
$H_2\times 10^4$ & 2.63 &0.04 & 0.01 & 0.02 & 0.04 & $^{+0.27}_{-0.13}$ & $^{+0.28}_{-0.15}$ \\ 
\hline
\end{tabular}
\end{center}
\caption {\label{table:h1h2}
Coefficients arising in the parametrization $H_P$ with their different
uncertainties.}
\end{table}
Here again, for practical purposes and for comparison with other
parametrizations of the vector form factor, it is useful to give the
first coefficients of the Taylor expansion  
\begin{eqnarray}
m_\pi^2 \left.\frac{d \bar{f}_+}{dt}\right|_{t=0} &=& \Lambda_+~, \nonumber \\
m_\pi^4 \left.\frac{d^2 \bar{f}_+}{dt^2}\right|_{t=0} &=& \Lambda_+^2 +
2 m_\pi^2 H'(0) = \Lambda_+^2 + (5.79^{+1.91}_{-0.97}) \times 10^{-4}~,\nonumber \\
m_\pi^6 \left.\frac{d^3 \bar{f}_+}{dt^3}\right|_{t=0} &=& \Lambda_+^3 +
6 m_\pi^2 H'(0) \Lambda_+ + 3 m_\pi^4 H''(0)  = \Lambda_+^3 + 3~(5.79^{+1.91}_{-0.97}) \times 10^{-4}\,\Lambda_+ 
\nonumber \\ &&+ (2.99^{+0.39}_{-0.21})\times 10^{-5}~.
\end{eqnarray}
The third order Taylor expansion is very accurate in this case,
too. The maximal error with respect to the exact dispersive representation is
$6\%$. 

\subsection{Calculation of $I_K$}
In order to extract $|f_+(0) V_{us}|$ from the measurement of the $K_{\ell 3}$ decay rate
\begin{equation}
\Gamma_{K^{+/0}_{\ell 3}}=\mathcal{N}_{K^{+/0}}~S_{EW}~(1+ 2\Delta^{EM}_{K^{+/0}\ell})~|f_+^{K^{+/0}}(0)V_{us}|^2~I^\ell_{K^{+/0}}~,~\mathcal{N}_{K^{+/0}}=C_{K^{+/0}}^2~G_F^2~m^5_{K^{+/0}}/(192 \pi^3)~,
\label{DecayW}
\end{equation}
one has to evaluate the phase space integrals $I^\ell_{K^{+/0}}$
defined in terms of the scalar and vector form factors:
\begin{equation}
I_{K^{+/0}}^\ell=\int_{m^2_\ell}^{t_0}~dt~\frac{1}{m_{K^{+/0}}^8}~\lambda^{3/2}~\left(1+\frac{m^2_\ell}{2t}\right)~\left(1-\frac{m^2_\ell}{2t}\right)^2~
\left(\bar f^2_+(t)+\frac{3 m^2_\ell \Delta_{K\pi}^2}{(2t+m^2_\ell)\lambda}\bar f_0^2(t)\right)~.
\label{IK}
\end{equation} 

Using the dispersive parametrization for the form factors, see Eqs.~(\ref{Dispf}, \ref{Dispfp}), it is possible
for practical purpose, to approximate the phase space integrals, Eq.~(\ref{IK}), by a polynomial expansion 
in terms of the two parameters ln$C$ and $\Lambda_+$ entering this parametrization. One obtains
\begin{equation}
I_{K^{+/0}}^\ell=c_0 + c_1 \Lambda_+ + c_2 \Lambda_+^2 + c_3 \Lambda_+^3 + c_4 \Lambda_+^4 + 
c_5 \mathrm{ln}C + c_6 \mathrm{ln}C^2 + c_7 \mathrm{ln}C^3 + c_8 \mathrm{ln}C^4~,
\label{IKappr}
\end{equation} 
where the polynomial coefficients for the 4 phase-space integrals are collected in the following table:
\begin{table} [h]
\begin{tabular}
[c]{|c|c|cccc|cccc|}\hline
& $c_{0}$ & $c_{1}$ & $c_{2}$ & $c_{3}$ & $c_{4}$ & $c_{5}$ & $c_{6}$ & 
$c_{7}$ & $c_{8}$ \\\hline
$I_{K^0}^e$ & 0.14126 & 0.48960 & 1.35655 & 3.12372 & 6.14597 & -- & -- & -- & -- \\
$I_{K^0}^\mu$ & 0.09061 & 0.29599 & 0.95764 & 2.36194 & 4.81855 & 0.01724 & 0.00515 & 0.00120 & 0.00023 \\
\hline
$I_{K^+}^e$ & 0.14530 & 0.53899 & 1.59964 & 3.94735 & 8.32532 & -- & -- & -- & -- \\
$I_{K^+}^\mu$ & 0.09324 & 0.32606 & 1.13005 & 2.98682 & 6.53170 & 0.01798 &
0.00545 & 0.00129 & 0.00025 \\ \hline
\end{tabular}
\caption{\it Coefficients of  $\Lambda_+$ and $\ln C$, Eq.~(\ref{IKappr})
in the polynomial expansion of 
the phase space integrals, Eq.~(\ref{IK}).} 
\end{table}


\end{appendix}


\begin{thebibliography}{99}
\bibitem{ISTRA}
  O.~P.~Yushchenko {\it et al.},
  Phys.\ Lett.\  B {\bf 581} (2004) 31
  [arXiv:hep-ex/0312004];
  O.~P.~Yushchenko {\it et al.},
  Phys.\ Lett.\  B {\bf 589} (2004) 111
  [arXiv:hep-ex/0404030].
\bibitem{KLOEe}
  F.~Ambrosino {\it et al.}  [KLOE Collaboration],
  Phys.\ Lett.\  B {\bf 636} (2006) 166
  [arXiv:hep-ex/0601038].
\bibitem{KLOEmu}
  F.~Ambrosino {\it et al.}  [KLOE Collaboration],
  JHEP {\bf 0712} (2007) 105
  [arXiv:0710.4470 [hep-ex]].
\bibitem{KTeVe}
  T.~Alexopoulos {\it et al.}  [KTeV Collaboration],
  Phys.\ Rev.\  D {\bf 70} (2004) 092007
  [arXiv:hep-ex/0406003].
\bibitem{KTeVz}
 E.~Abouzaid {\it et al.}  [KTeV Collaboration],
  Phys.\ Rev.\  D {\bf 74}, 097101 (2006)
  [arXiv:hep-ex/0608058]. 
\bibitem{NA48e} A.~Lai {\it et al.}  
  A.~Lai {\it et al.}  [NA48 Collaboration],
  Phys.\ Lett.\  B {\bf 604} (2004) 1
  [arXiv:hep-ex/0410065].
\bibitem{NA48mu}
  A.~Lai {\it et al.}  [NA48 Collaboration],
Phys.\ Lett.\  B {\bf 647} (2007) 341 [arXiv:hep-ex/0703002].
\bibitem{flavianet}
  M.~Antonelli {\it et al.}  [FlaviaNet Working Group on Kaon Decays],
  arXiv:0801.1817 [hep-ph].
\bibitem{GL84}
  J.~Gasser and H.~Leutwyler,
  Nucl.\ Phys.\  B {\bf 250} (1985) 517.
\bibitem{KN08}
  A.~Kastner and H.~Neufeld,
   Eur.\ Phys.\ J.\  C {\bf 57} (2008) 541
  [arXiv:0805.2222 [hep-ph]].
\bibitem{Bernard:06}
  V.~Bernard, M.~Oertel, E.~Passemar and J.~Stern,
  Phys.\ Lett.\  B {\bf 638} (2006) 480
  [arXiv:hep-ph/0603202].
\bibitem{hill}
  R.~J.~Hill,
  Phys.\ Rev.\  D {\bf 74}, 096006 (2006)
  [arXiv:hep-ph/0607108].
\bibitem{KTeVmu} V. Bernard, M. Oertel, E. Passemar, J. Stern and the KTeV collaboration, in preparation.  
\bibitem{Dashen:1969bh}
C. G. Callan and S. B. Treiman, Phys. Rev. Lett. {\bf 16} (1966) 153; 
  R.~F.~Dashen and M.~Weinstein,
  Phys.\ Rev.\ Lett.\  {\bf 22} (1969) 1337.
 \bibitem{Gasser:1984ux}
J.~Gasser and H.~Leutwyler,
Nucl.\ Phys.\ B {\bf 250} (1985) 517.
\bibitem{Leutwyler:07}
H. Leutwyler, private communication.  
\bibitem{Bijnens:07}
  J.~Bijnens and K.~Ghorbani,
  arXiv:0711.0148 [hep-ph].
\bibitem{BP08}
  V.~Bernard and E.~Passemar,
  Phys.\ Lett.\  B {\bf 661} (2008) 95
  [arXiv:0711.3450 [hep-ph]].  
\bibitem{Lellouch}
L. Lellouch, "Kaon Physics: A Lattice Perspective", Talk given at 
the XXVI International Symposium on Lattice Field Theory, 
July 14–19 2008, JLab, Virginia, USA, arXiv:0902.4545 [hep-lat].
\bibitem{BM}
E.~Blucher and W.~Marciano in Ref.~\cite{PDG:08}.
\bibitem{PDG:08}
C.~Amsler {\it et al.}  [Particle Data Group],
  Physics Letters B{\bf 667} (2008) 1.
\bibitem{Towner:08} 
  J.~C.~Hardy and I.~S.~Towner,
  arXiv:0812.1202 [nucl-ex].
\bibitem{omnes} 
R.~Omn\`es, 
Nuovo Cim.\  {\bf 8} (1958) 316;
N.~I.~Muskhelishvili, Singular Integral Equations, Noordhoff Series of Monographs on 
Pure and Applied Mathematics, Groningen (1953).
\bibitem{Donoghue}
  J.~F.~Donoghue, J.~Gasser and H.~Leutwyler,
  Nucl.\ Phys.\  B {\bf 343} (1990) 341.
\bibitem{jop00} M. Jamin, J. A. Oller and A. Pich,
Nucl. Phys. B {\bf 587} (2000) 279 [hep-ph/0006045]; ibid, 
Nucl. Phys. B {\bf 622} (2002) 279 [hep-ph/0110193]. 
\bibitem{Watson}
  K.~M.~Watson,
  Phys.\ Rev.\  {\bf 88} (1952) 1163.
\bibitem{Buettiker:2003pp}
P.~B\"uttiker, S.~Descotes-Genon and B.~Moussallam,
Eur.\ Phys.\ J.\ C {\bf 33} (2004) 409 [arXiv:hep-ph/0310283].
\bibitem{Estabrooks}
  P.~Estabrooks, R.~K.~Carnegie, A.~D.~Martin, W.~M.~Dunwoodie, T.~A.~Lasinski and D.~W.~G.~Leith,
  Nucl.\ Phys.\  B {\bf 133} (1978) 490.
\bibitem{Lepage}
G.~P.~Lepage and S.~J.~Brodsky,
Phys.\ Lett.\ B {\bf 87} (1979) 359.
\bibitem{Belle:07}
  D.~Epifanov {\it et al.}  [Belle Collaboration],
  Phys.\ Lett.\  B {\bf 654} (2007) 65
  [arXiv:0706.2231 [hep-ex]].
\bibitem{Babar:07}
  B.~Aubert {\it et al.}  [BABAR Collaboration],
  Phys.\ Rev.\  D {\bf 76} (2007) 051104
  [arXiv:0707.2922 [hep-ex]]; 
B. Aubert et al., talk presented at ICHEP08, Philadelphia, Pennsylvania, [arXiv:0808.1121 [hep-ex]].
\bibitem{Jamin:2008qg}
  M.~Jamin, A.~Pich and J.~Portoles,
  Phys.\ Lett.\  B {\bf 640} (2006) 176
  [arXiv:hep-ph/0605096];
ibid,
Phys.\ Lett.\  B {\bf 664} (2008) 78
[arXiv:0803.1786 [hep-ph]].
\bibitem{Boito:2008fq}
  D.~R.~Boito, R.~Escribano and M.~Jamin,
 Eur.\ Phys.\ J.\  C {\bf 59} (2009) 821
  [arXiv:0807.4883 [hep-ph]].
\bibitem{Bachir08}
  B.~Moussallam,
  Eur.\ Phys.\ J.\  C {\bf 53} (2008) 401
  [arXiv:0710.0548 [hep-ph]].
\bibitem{Aston:1987ir}
D.~Aston {\it et al.},
Nucl.\ Phys.\ B {\bf 296} (1988) 493.
\bibitem{Gounaris:1968mw}
  G.~J.~Gounaris and J.~J.~Sakurai,
  Phys.\ Rev.\ Lett.\  {\bf 21} (1968) 244.
\bibitem{Bernard:1990kw}
  V.~Bernard, N.~Kaiser and U.-G.~Mei{\ss}ner,
  Nucl.\ Phys.\  B {\bf 357} (1991) 129.
\bibitem{roessl99}
  A.~Roessl, 
Nucl.\ Phys.\ B {\bf 555} (1999) 507
  [arXiv:hep-ph/9904230].
\bibitem{Bern:04}  B.~Ananthanarayan, I.~Caprini, G.~Colangelo, J.~Gasser and 
H.~Leutwyler,  
Phys.\ Lett.\  B {\bf 602} (2004) 218  [arXiv:hep-ph/0409222].
\bibitem{Oller:07}  J.~A.~Oller and L.~Roca,
Phys.\ Lett.\  B {\bf 651} (2007) 139  [arXiv:0704.0039 [hep-ph]].
\bibitem{Gasser:2007de}
J.~Gasser,
PoS {\bf KAON} (2008) 033
[arXiv:0710.3048 [hep-ph]];
G.~Colangelo, J.~Gasser and A.~Rusetsky,
Eur.\ Phys.\ J.\  C {\bf 2009} (2009) 1 
[arXiv:0811.0775 [hep-ph]].
 
\bibitem{Kubis:2001ij}
  B.~Kubis and U.-G.~Mei{\ss}ner,
  Nucl.\ Phys.\  A {\bf 699} (2002) 709
  [arXiv:hep-ph/0107199].
\bibitem{Nehme:2001wf}
  A.~Nehme and P.~Talavera,
  Phys.\ Rev.\  D {\bf 65} (2002) 054023
  [arXiv:hep-ph/0107299].
\bibitem{BKOP08}
V. Bernard, M. Knecht, M. Oertel, E. Passemar, work in progress.  
\bibitem{leut02} H. Leutwyler,  in {\it Continuous Advances in QCD 2002}, eds.
K~A.~Olive, M.~A.Shifman and M.~B.~Voloshin,
World Scientific, Singapore (2003), p. 23, [hep-ph/0212324].
\bibitem{rasz76} I. Raszillier, W. Schmidt and I.-Sabba Stefanescu,
Z. Phys. A {\bf 277} (1976) 211.
\bibitem{oehme66}
R. Oehme, Phys. Rev. Lett. 16 (1966) 215.
\bibitem{hs04} J. Hirn and J. Stern, 
Eur. Phys. J. C {\bf 34} (2004) 447 [hep-ph/0401032];
ibid, JHEP {\bf 0409} (2004) 058 [hep-ph/0403017];
ibid, Phys.Rev. D {\bf 73} (2006) 056001 [arXiv:hep-ph/0504277].
\bibitem{stern06}
  J.~Stern,
  Nucl.\ Phys.\ Proc.\ Suppl.\  {\bf 174} (2007) 109
  [arXiv:hep-ph/0611127].
\bibitem{BOPS08}
  V.~Bernard, M.~Oertel, E.~Passemar and J.~Stern,
  JHEP {\bf 0801} (2008) 015
  [arXiv:0707.4194 [hep-ph]].
\bibitem{bij}
J.~Bijnens and P.~Talavera,
  Nucl.\ Phys.\  B {\bf 669} (2003) 341
  [arXiv:hep-ph/0303103].
\bibitem{jop04} M. Jamin, J. A. Oller and A. Pich,
JHEP {\bf 0402} (2004) 047 [arXiv:hep-ph/0401080].
\bibitem{ceekpp05} V.~Cirigliano, G.~Ecker, M.~Eidemuller, R.~Kaiser, A.~Pich and J.~Portoles,
JHEP {\bf 0504} (2005) 006 [arXiv:hep-ph/0503108]. 
\bibitem{Barate:99}
  R.~Barate {\it et al.}  [ALEPH Collaboration],
  Eur.\ Phys.\ J.\  C {\bf 11} (1999) 599
  [arXiv:hep-ex/9903015].
\bibitem{OPAL2}  G.~Abbiendi {\it et al.}  [OPAL Collaboration],
  Eur.\ Phys.\ J.\  C {\bf 35} (2004) 437
  [arXiv:hep-ex/0406007].

\end{thebibliography}
\end{document}